\documentclass[useAMS,usenatbib]{mn2e}

\usepackage{subfigure}
\usepackage{graphicx}
\usepackage[latin1]{inputenc}
\usepackage{amssymb}
\usepackage{natbib}

\usepackage{amsmath}
\bibpunct{(}{)}{;}{a}{}{,}
\newcommand{\apj}{ ApJ}
\newcommand{\aj}{ AJ}
\newcommand{\apjl}{ ApJL }
\newcommand{\mnras}{ M.N.R.A.S.}
\newcommand{\aap}{ Astron. Astrophys.}
\newcommand{\apjs}{ ApJS}
\newcommand{\araa}{ ARAA}
\newcommand{\HI}{H\,\textsc{i}}
\newcommand{\degree}{\ensuremath{^\circ}}
\def\ltsima{$\; \buildrel < \over \sim \;$}
\def\lta{\lower.5ex\hbox{\ltsima}}
\def\gtsima{$\; \buildrel > \over \sim \;$}
\def\simgt{\lower.5ex\hbox{\gtsima}}

\title[ISM \& Star Formation Demographics]{On the Interstellar Medium  and Star Formation Demographics of Galaxies in the Local Universe}
\author[M. S. Bothwell, R. C. Kennicutt, \& J. C. Lee]
{Matthew S. Bothwell$^{1}$\thanks{E-mail:
bothwell@ast.cam.ac.uk},
Robert C. Kennicutt, Jr$^{1}$ and
Janice C. Lee$^{2,3}$  \\
$^{1}$ Institute of Astronomy, University of Cambridge, Cambridge, CB3 0HA\\
$^{2}$ Carnegie Observatories, 813 Santa Barbara Street, Pasadena CA 91101\\
$^3$ Hubble Fellow}
\begin{document}
\date{Accepted 2009 July 30. Received 2009 July 17; in original form 2009 May 21}

\pagerange{\pageref{firstpage}--\pageref{lastpage}} \pubyear{2009}

\maketitle
\begin{abstract}
We present a demographic analysis of integrated star formation and gas properties for a sample of galaxies representative of the overall population at z $\sim$ 0. This research was undertaken in order to characterise the nature of star formation and interstellar medium (ISM) behaviour in the local universe, and test the extent to which global star formation rates can be seen as dependent on the interstellar gas content. Archival 21 cm derived \HI\ data are compiled from the literature, and are combined with  CO (J = 1 $\rightarrow$ 0) derived H$_2$ masses to calculate and characterise the \textit{total} gas content for a large sample of local galaxies. The distribution in stellar mass-normalised \HI\ content is found to exhibit the noted characteristic transition at stellar masses of $\sim 3 \times 10^{10} \;\mathrm{M}_{\sun}$, turning off towards low values, but no such transition is observed in the equivalent distribution of molecular gas. H$\alpha$ based star formation rates (SFRs) and specific star formation rates (SSFRs) are also compiled for a large (1110) sample of local galaxies. We confirm two transitions as found in previous work: a turnover towards low SFRs at high luminosities, indicative of the quenching of SF characteristic of the red sequence; and a broadening of the SF distribution in low-luminosity dwarf galaxies, again to extremely low SFRs of $< 10^{-3} \;\mathrm{M}_{\sun}\; \mathrm{yr}^{-1}$. However, a new finding is that while the upper luminosity transition is mirrored by the turn over in  \HI\ content, suggesting that the low SFRs of the red sequence result from a lack of available gas supply, the transition towards a large spread of SFRs in the least luminous dwarf galaxies is \textit{not} matched by a prominent increase in scatter in gas content.  Possible mass-dependent quenching mechanisms are discussed, along with speculations that in low mass galaxies,  the H$\alpha$ luminosity may not faithfully trace the SFR.
\end{abstract}

\section{Introduction}
The formation of stars is one of the primary drivers of galaxy evolution, and obtaining a full quantitative understanding of this process is an indispensable aspect of any investigation into galaxy behaviour. A profound symbiosis exists between the phenomenon of star formation and the form and content of the interstellar medium (ISM) of the host galaxy: a duel analysis, therefore, addressing both the star formation behaviour and gas distribution in a sample of galaxies is ideally suited to characterising the properties of local galaxies, and analysing the results in the context of more global trends. 

Recent investigations into the evolutionary properties of galaxies have resulted in the striking finding that the distribution of galaxy colours exhibits a marked bimodality, separating `red and dead' early type galaxies with little or no ongoing star formation, from a `blue sequence' of star forming spirals, with the divide occurring at M$_{\mathrm{B}} \sim -19$ (\citealt{2001AJ....122.1861S}, \citealt{2004ApJ...608..752B}, \citealt{2004ApJ...600..681B}, \citealt{2007ApJS..173..315S}). This bivariate distribution can be examined in terms of many other galactic properties, such as the SFR; the colour-magnitude diagram can be interpreted in terms of SF history and stellar mass content (e.g. \citealt{2004ApJ...613..898T}, \citealt{2004MNRAS.351.1151B}, \citealt{2007ApJ...660L..47N}, \citealt{2007tS..173..267S}).

The divide between the red and blue sequences emerges also in the star formation rate per unit stellar mass (Specific Star Formation Rate, SSFR), as demonstrated by \cite{2004MNRAS.351.1151B} and \cite{2007ApJS..173..315S}: the distribution in the SSFR vs. M$_*$ plane is bimodal, with a red sequence of low SFR massive galaxies (SSFR $\sim 10^{-12} \;\mathrm{yr}^{-1}$), and a less massive blue sequence of galaxies with significatly higher SSFRs ($10^{-9} - 10^{-10}\;\mathrm{yr}^{-1}$). \cite{2007ApJS..173..315S} use these areas to classify galaxies as either on or off a `SF sequence', 
along which many SF parameters (such as the SFR surface density, $\Sigma_{\mathrm{SFR}}$) are close to constant. In this picture, the blue `main sequence' galaxies evolve secularly towards higher M$_*$ while gradually lowering their SSFRs, while the red sequence turn over is caused by the quenching of the SF. Conversely, galaxies can be driven off the main sequence in the other direction (towards higher SSFRs) by an event - such as an interaction - that triggers an increased SFR, or a sudden starburst (e.g. \citealt{1989Natur.340..687H}; \citealt{1991ApJ...370L..65B}; \citealt {2006ApJS..163....1H}).

The rate at which a galaxy is forming stars is, however, in some sense a `high level process', being both complex and dependent on the more fundamental underlying properties of the galaxy - primarily the content of the ISM. The galactic `gas fraction' (as opposed to the absolute amount of gas, which is strongly dependent on the mass of the host galaxy) provides information on the evolutionary history of the system in question, and also determines the potential for future star formation. As such, an understanding of how the gas fraction varies from galaxy to galaxy may help shed light on the origin of the bifurcation in the colour-magnitude plane; tentative explorations of the causes underlying this phenomena have highlighted the role of gas - \cite{2006MNRAS.368....2D}, for example, note that the bifurcation mass coincides with the transition between the `cold accretion' and `cooling flow' regimes of model galaxy formation. However, the role of interstellar gas in the development of the range of galaxies we see today has not been fully explored. A key question to address is whether the bimodal division into two galactic types can ultimately be seen as a consequence of the underlying gas behaviour.


Investigations into the global properties of galaxies have principally focused on the upper end of the luminosity function, due to the ease at which brighter systems can be studied. However, investigations into star forming behaviour in low mass dwarf galaxies have resulted in some unexpected findings. A study by \cite{2007ApJ...671L.113L} investigated the distribution of the SSFR (measured via the H$\alpha$ Equivalent Width) with absolute magnitude for a small volume limited sample of local galaxies (obtained from the 11Mpc H$\alpha$ UV Galaxy Survey, [11HUGS]), with H$\alpha$ derived SFRs for the population of local low luminosity dwarf galaxies, down to M$_{\mathrm{B}} \sim -10$. In addition to the well studied transition (found by \cite{2007ApJ...671L.113L} to be at M$_{\mathrm{B}} = -19$), a second transition was found, with low mass dwarf galaxies (M$_{\mathrm{B}} >$ -15) having a very large ($>$ 3 dex) spread in their specific star formation rate (SSFR). Between the two transition regions, the SSFR was found to inhabit a very restricted range of values, implying a high degree of regulation in the star forming behaviour of the blue sequence - this comprises the `main sequence' of star forming galaxies discussed above. These findings suggest that there may be three distinct modes of star forming behaviour operating across the mass spectrum, rather than two as previously thought. A lack of correlation with environmental factors \citep{2004AJ....128.2170H} suggests that internal processes are responsible for this third mode, but the nature of these process is, as yet, unclear.

In this paper, the star formation properties and interstellar medium (ISM) content of a sample of local galaxies are characterised in order to delineate the various populations that exist along the low-redshift Hubble sequence. As such, we seek to determine the extent to which the previously described bimodality in the SSFR is solely a result of the standard scaling of star formation rate with gas content  (e.g. \citealt{1959ApJ...129..243S}; \citealt{1998ApJ...498..541K}). 

Our primary sample, the 11HUGS dataset, is highly complete at the lower end of the mass function, but suffers from a paucity of bright massive objects: while over 80\% of the sample has a SFR lower than the LMC, just $\sim$6\% have SFRs greater than the Milky Way (taking SFR$_{\mathrm{MW}}$=2-6 M$_{\sun}$ yr$^{-1}$, as reported by \citealt{1999MNRAS.307..857B}). Therefore, we aim to augment the 11HUGS dataset with more voluminous surveys, to see whether the inclusion of more data at the brighter end of the luminosity function would corroborate previous findings. Our overall compiled sample, consisting of nearly a thousand local galaxies, provides a dataset with not only excellent completeness at low ($\sim$SMC) masses, but also coverage over a large enough volume to statistically sample galaxies with masses comparable to or greater than the Milky Way. This sample allows star formation rates and gas contents to be compared in a uniform manner over a mass spectrum spanning five orders of magnitude. This paper will also re-evaluate the possibility of a `third mode' of star formation existing in low mass dwarfs, again in the context of the available gas supply.

In \S 2, we describe the samples and data drawn upon for this work, and briefly describe the calculations we have used to derive various parameters. In \S 3, a demographic analysis of the gas behaviour, star forming behaviour, and the link between the two are presented in turn for both a highly complete local volume sample, and a more populous deeper sample. These results are discussed in more detail in \S 4, and we present conclusions in \S 5.

\section[]{Methodology}
\subsection{Sample Description}
\label{sec:samples}
There were two primary drivers behind the selection of data. Firstly, a large sample was required, in order to represent as fully as possible the trends in gas properties and star forming behaviour in the local universe. In order for this aim to be realised, the sample must be relatively homogeneous, in order to investigate the range of `normal'  behaviours; many galaxies are observed precisely because of their unusual behaviour, with the result that an uncritical compilation of H$\alpha$ and \HI\ data from the literature would likely result in biases towards more extreme objects. Secondly, in order to characterise the full range of behaviours across the entire mass spectrum, good completeness at the lower end of the luminosity function was required. Low mass dwarf galaxies are the numerically dominant galaxy type, but Malmquist-type biases result in their exclusion from many observational investigations. As a result, volume limited local surveys (which will be dwarf-dominated) have been utilised. The resulting combined sample has the advantage of being highly complete at low masses, and large enough to be representative at high masses. In order to ensure that the final compilation is as bias-free as possible, the analysis has been repeated for the local volume sample (see \S3.2), with no change in result. 

\subsection{H$\alpha$}
Information on the star formation properties of galaxies in the Local Volume was obtained from the H$\alpha$ survey of \cite{2008ApJS..178..247K}. This survey is an essentially volume limited sample of 436 local galaxies (D $\leq$ 11 Mpc). Galaxies were selected with the added proviso that galaxies in the so called `zone of avoidance' (e.g. lying in the galactic plane, $|b| \geq 20\degree$) were excluded. This survey includes the 436 known galaxies within 11 Mpc, and is essentially complete for M$_{\mathrm{B}} \sim $ -15.5, and M(\HI) $\sim 10^8$ M$_{\sun}$ \citep{2009ApJ...692.1305L}, providing excellent statistics on the star formation rates of the low luminosity dwarf population. 


This sample has good statistical sampling of the lower end of the luminosity function, but due to the rarity of bright (M$_{\mathrm{B}} <$ -19) galaxies in the local volume surveys covering larger volumes were required to achieve improved statistical coverage at the upper end. The survey of \cite{2004A&A...414...23J} was chosen for this purpose. This is a H$\alpha$ based survey of 334 galaxies, taken from the Uppsala General Catalogue (UGC), which has a magnitude limit of B = 15.5. The galaxies examined in the James et al. H$\alpha$ survey were chosen to have recession velocities of $<$ 3000 km s$^{-1}$ (D$\lta$43 Mpc; assuming H$_0$ = 70 km s$^{-1}$Mpc$^{-1}$). James et al. also stipulated that the galaxies have a morphological T-type $\geq$ 0 (S0/a or later). 

In addition to the sample of \cite{2004A&A...414...23J}, many galaxies were included from the large database of H$\alpha$ galaxies (Kennicutt et al. 2010, in prep).  More than 3000 galaxies have been observed in H$\alpha$ \citep{2008ApJS..178..247K}, providing a rich and varied dataset to sample. However, the data are extremely heterogeneous, with many of the galaxies being investigated precisely because of their unusual, or indeed exceptional, behaviour. In order to provide a characteristic representation of local universe star formation, unbiased towards extreme objects, surveys and compilations specifically targeting cluster members (such as the GOLDMine sample of \citealt{2003A&A...400..451G}) and starburst galaxies were excluded.
The complete H$\alpha$ sample consists of 1110 galaxies.

\subsection{HI}

Detection of the neutral atomic component of the ISM in low redshift galaxies is comparatively easy, due to the high sensitivity available at wavelengths of 21cm \citep{1994ARA&A..32..115R}, and as such many large \HI\ surveys exist in the literature. For this study, information on the neutral hydrogen content of galaxies was obtained from the large homogeneous compilation presented by \cite{2005ApJS..160..149S}. This is a large ($\sim$ 9000) galaxy compilation of optically selected galaxies in the low redshift universe, collected over a period of 20+ years, selected to have heliocentric velocities of $-200 < \mathrm{V}_{\sun} < 28,000 \;\mathrm{km s}^{-1}$. A total of 8850 galaxies were detected unequivocally in \HI\, with a further 156 and 494 being marginally detected and not detected respectively. 

In order to remove the strong effect of galactic stellar mass on the \HI\ content, and to provide a parameter as closely analogous to SSFR as possible, the \HI\ content is presented in the form of a \HI\ mass to stellar mass ratio, \textit{R}(\HI), defined as M$_{\mathrm{HI}}$/M$_*$. This constrains the analysis to those galaxies with available photometry in order to measure the stellar mass (see \S 2.5 below). A total of  7486 `detected' galaxies from the Springob survey possess archival B-band photometry - this makes up the bulk of the \HI\ sample.


In addition, to provide \HI\ information for local dwarf galaxies, the \HI\ data from the \cite{2004AJ....127.2031K} compilation will be included in the analysis. This is an all sky catalogue of 451 galaxies, selected to have a distance $\leq$ 10 Mpc ($= \mathrm{V}_{\mathrm{rad}}$ $\leq$ 550 km s$^{-1}$), which contains both optical photometry and 21cm \HI\ data, making it ideal for an analysis of ISM content. The limiting magnitude of the survey was B$_t$ = 17.5, with a resulting absolute magnitude limit of M$_{\mathrm{B}} \sim$ -7 at 1 Mpc, and M$_{\mathrm{B}} \sim$ -12 at 10 Mpc. 

Additional 21 cm \HI\ data to complement the H$\alpha$ dataset was obtained from the sample of Springob et al., the NASA Extragalactic Database (NED), and the online reference catalogue of astrophysical data  HYPERLEDA (\citealt{2003A&A...412...45P}).

The total number of galaxies included in the \HI\ analysis is therefore 7819.

\begin{figure}
 \centerline{\includegraphics[scale=0.55]{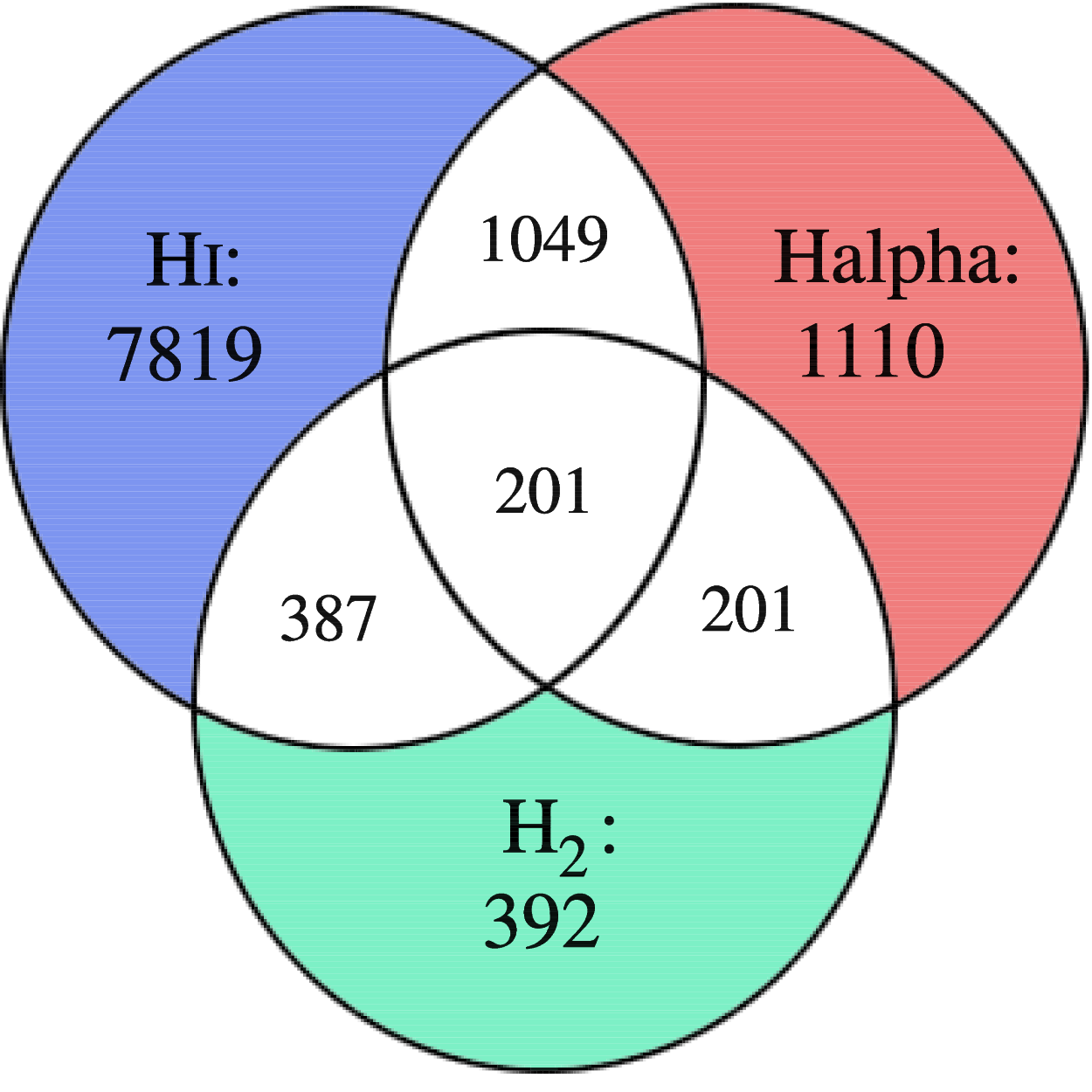}}
 \caption{Schematic Venn diagram illustrating the 3 data types, respective sample sizes, and number of galaxies with combinations of data. Note that the numbers in the coloured segments represent the \textit{total} number of galaxies recorded with the respective data type, as opposed to the standard Venn formalism whereby the numbers represent galaxies recorded \textit{exclusively} in the respective data type.}
 \label{fig:venn}
\end{figure}

\subsection{H$_2$}
The study of molecular hydrogen is important to the understanding of star formation: to the best of our knowledge, it is only inside the giant molecular cloud (GMC) complexes that stars can form. H$_2$ is much more challenging to detect than \HI; few galaxies have been mapped in CO (the FCRAO extragalactic survey of \cite{1995ApJS...98..219Y}, for example, contains resolved CO data for 300 galaxies), and fewer still have the unresolved CO measurements required to measure the global CO flux, necessary to calculate a \textit{total} molecular hydrogen mass. 229 galaxies in the FCRAO survey have global integrated CO fluxes recorded. In addition, archival CO data for 245 galaxies has been compiled and presented by \cite{2009arXiv0901.2526O}, which were added to the sample. H$_2$ is very difficult to detect in dwarf galaxies; they have little to no CO emission, and uncertainties in the CO-H$_2$ conversion factor at low metallicities mean that it is extremely challenging to get accurate H$_2$ masses for the lowest luminosity galaxies in our sample. A recent survey by \cite{2005ApJ...625..763L} attempts to address this issue, and from there we take 10 sources that overlap with our sample. A total of 392 unique galaxies with single dish CO fluxes were used, 43 of which have only upper limits on their CO fluxes.

Figure \ref{fig:venn} shows a diagrammatical representation of the sample sizes for different data types, and the number of sources with various combinations of available data.


\subsection{Derived Parameters}

\begin{flushleft}
\textbf{H\textsc{I} mass}: 
\end{flushleft}
\HI, detected via the hyperfine splitting spin transition of atomic hydrogen emitting at 21 cm ($\nu$ = 1420 MHz), is the most powerful available probe of the ISM. The forbidden nature of the transition results in most accumulations of \HI\ being optically thin, with a direct proportionality between the 21 cm luminosity and the number of emitting atoms. The neutral hydrogen mass is therefore given by the simple formula
\[
\mathrm{M}_{\mathrm{HI}} = 2.36 \times 10^5 \;\mathrm{D}^2 \int Sdv \;\;\; \mathrm{M}_{\sun} 
\]
Where $\int Sdv$ is the integrated 21cm line flux in Jy km s$^{-1}$, and the distance D is in Mpc.

\begin{flushleft}
\textbf{H$_2$ mass}:
\end{flushleft}
The molecule H$_2$ lacks a permanent electric dipole moment, and as such is very difficult to detect directly. Instead, CO emission (e.g.\ the J = $1 \rightarrow 0$ transition at $\lambda = 2.6$mm) is measured, and converted to a H$_2$ column density (e.g. \citealt{1995ApJS...98..219Y})
The `standard' calibration used (the oft discussed CO-to-H$_2$ conversion factor, given by \citealt{1997ASSL..161...33K}) 
 is 
 \[
 \mathrm{N}_{\mathrm{H}_2} / \mathrm{I}_{\mathrm{CO}} = \mathrm{X}_{\mathrm{CO}} \times 10^{20} \;\mathrm{H}_2\; \mathrm{cm}^{-2} \;[\mathrm{K}(\mathrm{T}_{\mathrm{R}}) \;\mathrm{km\, s}^{-1}]^{-1}
 \]
Where $\mathrm{X}_{\mathrm{CO}}$ is generally taken to be 2 - 3; we take the X-factor to be a constant value of 2.8. Kenney and Young (1989) show that the molecular hydrogen mass can therefore given by
\[
\mathrm{M}_{\mathrm{H}_2} = 1.1\times 10^4  \cdot \;\mathrm{D}^2 \;\mathrm{S}_{\mathrm{CO}} \;\;\; \mathrm{M}_{\sun} 
\]
Where S$_{\mathrm{CO}}$ is the global  J = ($1 \rightarrow 0$) flux in Jy km s$^{-1}$, and the distance D is in Mpc. This assumption of a constant CO-to-H$_2$ conversion factor (as opposed to, say, a luminosity dependent factor as presented by \citealt{2009arXiv0901.2526O}) will have the tendency to underestimate the mass of molecular hydrogen in the low mass dwarf galaxies. However, the quantities involved are negligible fractions of the total mass of the ISM; for this analysis, therefore, the constant factor has been used. 

\begin{flushleft}
\textbf{Absolute Magnitudes}: 
\end{flushleft}
Absolute magnitudes (M$_{\mathrm{B}}$) were derived from the total B-band magnitudes corrected for internal extinction and inclination effects (B$_{\mathrm{T}}^0$) given in the RC3 of \cite{1991trcb.book.....D}, using the standard conversion formula

\[
\mathrm{M} = \mathrm{m} - 5 \log \mathrm{D}_{\mathrm{L}} + 5
\] 
Where corrected B-band magnitudes were unavailable, the formulae given by de Vaucouleurs et al. were used to convert total B magnitudes into this system:
\[
\mathrm{B}_{\mathrm{T}}^0 = \mathrm{B}_{\mathrm{T}} - \mathrm{A}_{\mathrm{g}} - \mathrm{A}_{\mathrm{i}} - \mathrm{K}_{\mathrm{B}}
\]
Where $\mathrm{B}_{\mathrm{T}}$ is the total B-band magnitude, $\mathrm{A}_{\mathrm{g}}$ is the galactic extinction (calculated here using the IR maps of \citealt{1998ApJ...500..525S}), $\mathrm{K}_{\mathrm{B}}$ is the K correction for redshift, and $\mathrm{A}_{\mathrm{i}}$ is a linear function of $\log \mathrm{R}_{25}$ (which here is the isophotal axis ratio, as opposed to an isophotally-defined radius),

\[
\mathrm{A}_{\mathrm{i}} = \alpha(\mathrm{T}) \log \mathrm{R}_{25}
\]
with
\[
\alpha(\mathrm{T}) = 0 \;\;\;\;\;\;\;\;\;\;\;\;\;\;\;\;\;\;\;\;\;\;\;\;\;\;\;\;\;\;\;\;\:\:  \mathrm{T} < 0
\]
\[
\alpha(\mathrm{T}) = 1.5 - 0.03(T - 5)^2 \;\;\;\;\;\;\; \mathrm{T} \geq 0
\]
The K-correction has been taken to be negligible, as the most distant galaxy is UGC 12840 at 93 Mpc ($z = 0.023$ for a concordance cosmology of $\Omega _\mathrm{m} = 0.3$,  $\Omega _\Lambda = 0.7$, $\mathrm{h} = 0.73$), and the vast majority are closer than 40 Mpc ($z < 0.01$).

\begin{flushleft}
\textbf{Stellar Mass}: 
\end{flushleft}
\cite{2001ApJ...550..212B} note that the stellar mass to light ratio shows trends with various galaxy parameters (including surface brightness, gas fraction, K-band magnitude, colour, and galaxy mass), and - most critically for this study - has a dependency on the recent star formation history of the galaxy in question. Using any given photometric luminosity (even the K-band) as a proxy for the stellar mass will therefore introduce errors, as there is no passband where the mass to light (M/L) ratio is constant. Bell \& de Jong's model (M/L) ratios correlate strongly with galactic colour; they therefore provide the algorithm 
\[
\log(\mathrm{M/L}) = a_{\lambda} + b_{\lambda} \cdot \mathrm{colour}
\]
where for B - V colours and B-band photometry, $a_{\lambda}$(B) = -1.110 and $b_{\lambda}$(B) = 2.018, for a scaled Salpeter IMF and a mass dependent formation epoch model (as described by, for example, \citealt{2000MNRAS.319..168C}).

\begin{figure}
 \centerline{\includegraphics[scale=0.55]{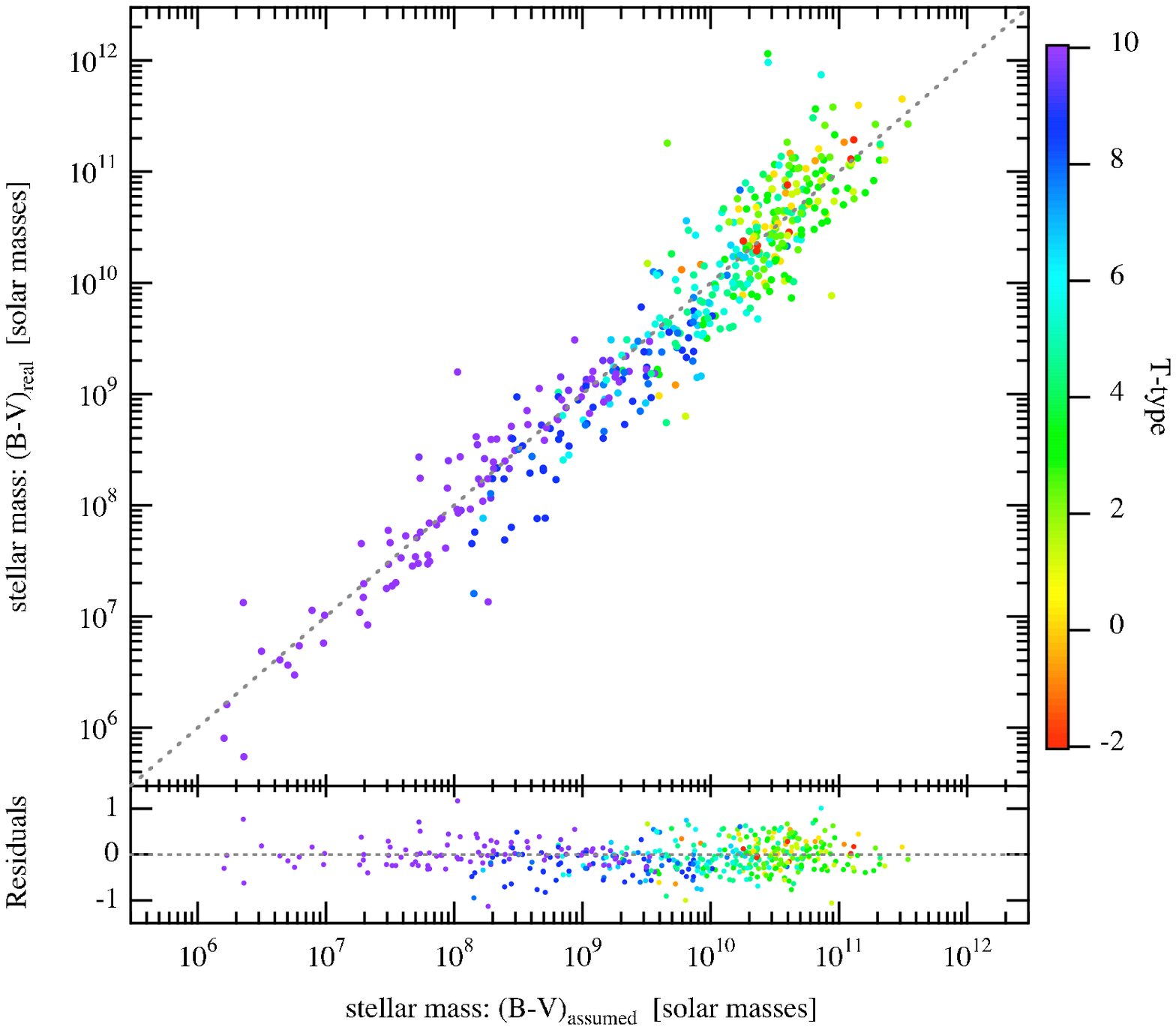}}
 \caption{Plot of the stellar mass, derived using the method of Bell and de Jong (2001). The y-axis displays the stellar mass derived using real B-V photometric colours, and the x-axis displays the stellar mass derived using the assumed B-V colour given in Table \ref{tab1}.}
 \label{fig:mass}
\end{figure}

A potential problem with this method is the lack of complete photometric datasets from which the colours can be derived: archival B - V photometry (corrected for reddening) is only available for $\sim 50\%$ of the sample (498 galaxies of the total 1049). However, it is possible to accurately estimate a colour based on morphological type (e.g. \citealt{1958MeLu2.136....1H}; \citealt{1994ARA&A..32..115R}), where the morphological types are coded in the RC3 system as reported in the Hyperleda database. The colours are taken from \cite{1994ARA&A..32..115R}, and are shown in Table \ref{tab1}. Figure \ref{fig:mass} shows the stellar mass derived from the real and assumed colours for the 498 galaxies with B - V photometry. It was found that using the colours derived solely from morphological type data caused the stellar mass to be significantly underestimated for the lowest mass dwarfs. To correct for this effect, galaxies with masses calculated to be $< \;10^{7.5} \;\mathrm{M}_{\sun}$ were assigned a (B-V) colour of 0.3 - this resolved the issue. The two methods for mass estimation agree well, with reasonably small scatter (see \S \ref{sec:err} for discussion) and no significant trend in the residuals: this suggests that the colour assumption method is a acceptable one for the estimation of stellar masses; this assumptive technique will be used to estimate stellar masses for the remaining 40\% of the sample for which no colour is available. It is important to note that restricting the analysis in this work to only those galaxies with archival colours (and, therefore, reliable stellar mass estimates) does not alter the scatter or trends in any of the derived relationships, and does not alter any of the conclusions drawn.

\begin{table}
\centering
\caption{Morphology-Colour Relationship, with resultant B-band mass-light ratio (M/L): colours from Roberts \& Haynes (1994), M/L ratios calculated using the algorithm of Bell \& de Jong (2001). $^*$Note that galaxies with stellar masses $< 10^{7.5}$ M$_{\sun}$ were assigned a (B-V) colour of 0.3 - see text for details.}
\begin{tabular}{ccc}
\hline
T-type & $<\mathrm{B} - \mathrm{V}>$  & M/L \\
 \hline
 \hline
S0 & 0.9 & 5.08 \\
S0a & 0.8 &3.19\\
Sa & 0.7&2.00\\
Sab & 0.7&2.00\\
Sb & 0.6&1.26\\
Sbc & 0.55&1.00\\
Sc & 0.5&0.79\\
Scd & 0.5&0.79\\
Sd & 0.5&0.79\\
Sm & 0.4&0.49\\
Im & 0.4&0.49\\
dIrr$^*$ & 0.4&0.49\\
\hline
\end{tabular}
\label{tab1}
\end{table}


\begin{flushleft}
\textbf{SFR}: 
\end{flushleft}
Photoionisation of interstellar gas by massive ($>$ 10 M$_{\sun}$) stars results in Balmer line emission, with a flux proportional to the number of incident ionising UV photons. By measuring the global flux in a Balmer emission line (notably H$\alpha$), it is possible to calculate the ionising UV flux, and therefore the number of O and B stars in a galaxy. O and B stars being very young, it is possible (assuming some IMF) to extrapolate to the total instantaneous star formation rate. \cite{1998ARA&A..36..189K} gives the conversion
\[
\mathrm{SFR \;\; (M_{\sun} \; year)}^{-1} = 7.9 \times 10^{-42} \mathrm{\; \;L(H\alpha) \;\;(ergs \, s}^{-1})
\]
Which is based on a standard Salpeter IMF ($\alpha = 2.35$) from 0.1 - 100 M$_{\sun}$. Most narrowband surveys for H$\alpha$ use filters which are wide enough to detect the adjacent [NII]6548,6583 lines, and simply publish the combined H$\alpha$+[NII] flux. In such cases, a correction for [NII] contamination must be applied (see \citealt{2008ApJS..178..247K}):
\[
\log (\mathrm{[NII]/H}\alpha)  = (-0.173 \pm 0.007) \mathrm{M}_{\mathrm{B}} - (3.903 \pm -0.137) 
\]
for $\mathrm{M}_{\mathrm{B}} > -21$, and 
\[
\log (\mathrm{[NII]/H}\alpha)  = 0.54 
\]
for $\mathrm{M}_{\mathrm{B}} \leq -21$.

\subsection{Nebular Attenuation Corrections}
The internal H$\alpha$ extinction is crucial to constrain in order to accurately determine star formation rates. Internal extinction due to dust is the largest source of systematic error when measuring SFRs \citep{1998ARA&A..36..189K}, and the large variations between galaxies can also introduce significant random errors. 

The simplest (and crudest) method of correcting for extinction is to adopt a constant extinction factor (1.1 magnitudes was used by \cite{1983ApJ...272...54K}, for example); this method is lacking in rigour, however, and is insufficient for a thorough analysis, as the real amount of extinction can vary widely between galaxies \citep{1996A&A...306...61B}. A method which takes into account the general trend that the attenuation increases with increasing galaxy luminosity (i.e. dwarf galaxies are more transparent on average than more massive spiral discs) is given in \cite{2009ApJ...692.1305L} as:

\[
\mathrm{A(H}\alpha) = 0.14 \;\;\;\;\;\;\;\;\;\;\;\;\;\:\;\;\;\;\;\;\;\;\;\;\;\;\;\;\;\;\;\;\;\;\;\;\;\;\;\;\;\;\;\;\; (\mathrm{M}_{\mathrm{B}} \geq -15)
\]
\[
\mathrm{A(H}\alpha) = 1.971+0.323\mathrm{M}_{\mathrm{B}}+0.0134\mathrm{M}_{\mathrm{B}}^2 \;\;\;\;\; (\mathrm{M}_{\mathrm{B}} < -15)
\]

This is calculated from the Balmer decrement, using the spectroscopic measurements of \cite{2006ApJS..164...81M}. A more rigourous method is to calculate the H$\alpha$ extinction directly from the total infrared (TIR) luminosity of a galaxy.  This method is based on an `energy balance' assumption that the extincted H$\alpha$ flux will be re-radiated in the IR (see \citealt{Kennicutt_prep_09}  for details). \cite{Kennicutt_prep_09} give the prescription 

\[
\mathrm{A(H}\alpha) = 2.5 \log\left(1 + 0.0025 \;\frac{\mathrm{L(TIR)}}{\mathrm{L(H\alpha)}_{\mathrm{obs}}}\right)
\]

for calculating A(H$\alpha$), the extinction in H$\alpha$. Due to the self-similar shape of the infrared spectral energy distribution (SED) at different luminosities, the total IR luminosity can be accurately estimated (to better than 1\% for most galaxies) using a superposition of 25$\mu$m, 60$\mu$m, and 100$\mu$m IRAS fluxes, using the algorithm of \cite{2002ApJ...576..159D};

\[
\mathrm{L(TIR)} = \zeta_1\nu\mathrm{L}_{\nu}(25\mu\mathrm{m})+\zeta_2\nu\mathrm{L}_{\nu}(60\mu\mathrm{m})+\zeta_3\nu\mathrm{L}_{\nu}(100\mu\mathrm{m})
\]

where $\zeta_1 =  2.403$, $\zeta_2 = -0.2454$, and $\zeta_3 = 1.6381$ at $z=0$. Estimating the extinction in this way does require that the galaxy in question has been observed by IRAS, which is only true of a small percentage of the total sample. If no IRAS fluxes are available, the extinction will be taken to be the quadratic function of M$_{\mathrm{B}}$ described above. 

%
%
%

\section{Analysis}

In order to investigate the properties of our sample of galaxies, a demographic analysis has been carried out on two characteristic parameters, i) \textit{R}(gas) and ii) the star formation rate (both of which are closely linked via the Schmidt law formalism), and their trends with absolute magnitude.  

\subsection{Gas Demographics}
In this analysis of the gas demographics, the atomic and molecular phases of the ISM are presented separately. First, the sample of  \cite{2005ApJS..160..149S} is combined with the Local Volume \HI\ data of \cite{2004AJ....127.2031K}, to provide a large compilation of \textit{R}(\HI) ($= \mathrm{M}_{\mathrm{HI}} / \mathrm{M}_*$). Secondly, the available H$_2$ data are presented in the same way. Finally, the \HI\ data are combined with H$_2$ data to present total gas ratios, \textit{R}(total).

\subsubsection{Atomic Hydrogen}
\begin{figure}
 \centerline{\includegraphics[scale=0.58]{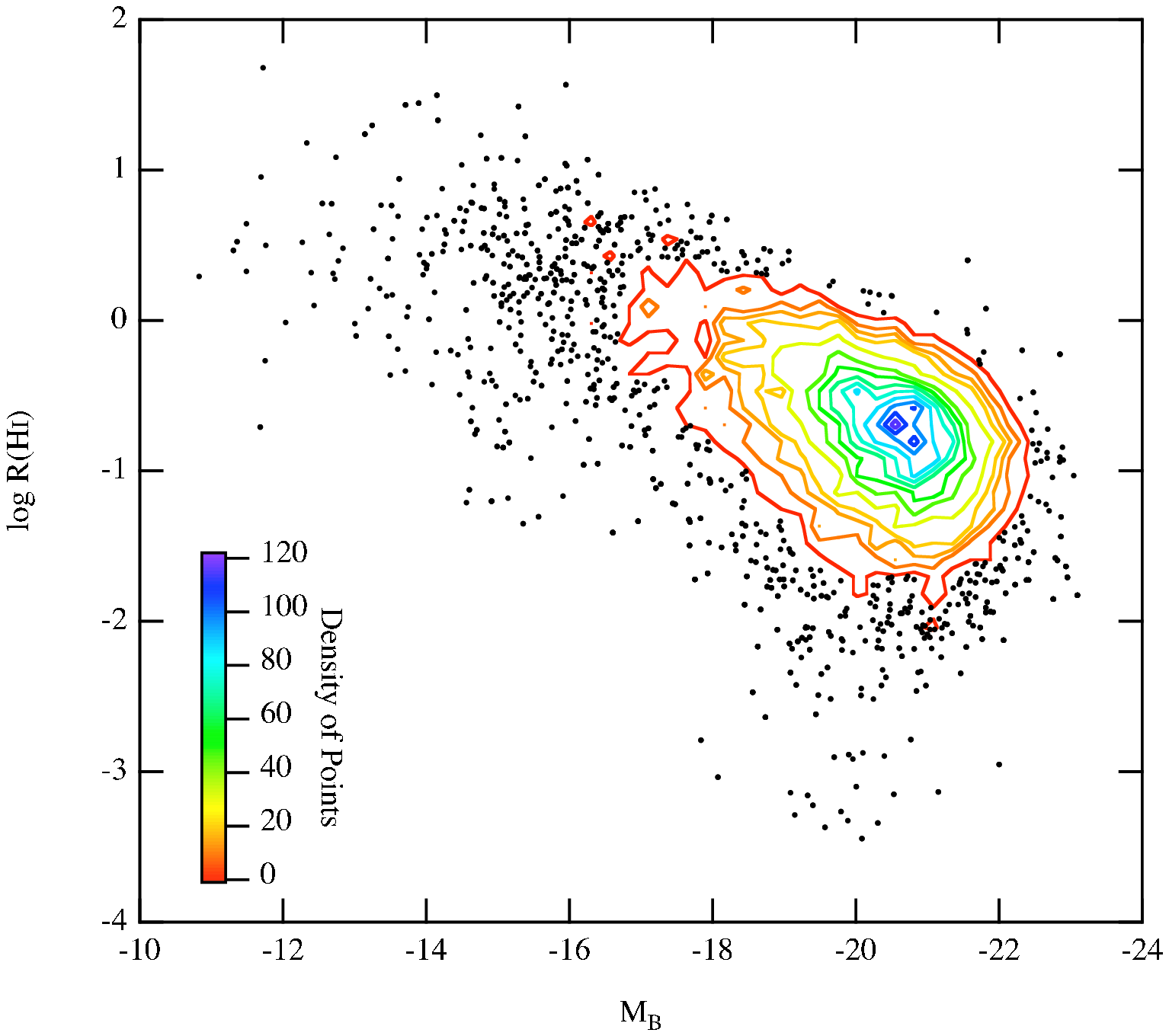}}
 \caption{\textit{R}(\HI) plotted against absolute B-band magnitude for the 7819 galaxies from the \HI\ sample with concomitant B-band photometry and morphological type data. Contours have been used where the density of points is high; where points become too rarified for contours to faithfully trace the distribution, individual points have been plotted. The Least-squares best fitting power law fitted to the distribution has an index of ($-0.58 \pm 0.007$).}
 \label{fig:hi}
\end{figure}

In order to remove the effect of galaxy mass on the \HI\ content, we normalise to the stellar mass, and present \textit{R}(\HI), defined as M$_{\mathrm{HI}}$/M$_{\mathrm{*}}$. Figure \ref{fig:hi} plots \textit{R}(\HI) against absolute B-band magnitude for the entire \HI\ sample as described above. 

Taking the Hubble sequence to be a rough function of luminosity, it is clear that Figure \ref{fig:hi} reproduces the (by now) well known result (e.g. \citealt{1994ARA&A..32..115R}) that the \HI\ ratio increases along the Hubble sequence, with the smaller (late type) galaxies typically having\textit{R}(\HI) $> 0.5$ (and most having \textit{R}(\HI) $> 1$), and the mean value of \textit{R}(\HI) decreasing as luminosity increases (and the population shifts to being made up of earlier types). The dense centre of the plot (i.e. the peak of the sample's number distribution) is dominated by intermediate type (Sa, Sb, Sc) discs, which have typical \textit{R}(\HI) of $\sim$ 0.2.

The distribution is smooth for M$_{\mathrm{B}} > -18 $, with the mean gas mass ratio decreasing monotonically with increasing luminosity. At M$_{\mathrm{B}} < -18$, however, the distribution is joined by outliers existing below the main locus, as early types appear with extremally low \textit{R}(\HI) of $< 0.05$. The entire distribution (in $\log$-\textit{R} - $\log$-L$_{\mathrm{B}}$ space) can be fit by a single power law of index $-0.58 \pm 0.007$ (see \S \ref{sec:ES} for a fuller discussion of the slopes of the distributions).

\subsubsection{Total Gas Content}

\begin{figure}
 \centerline{\includegraphics[scale=0.7]{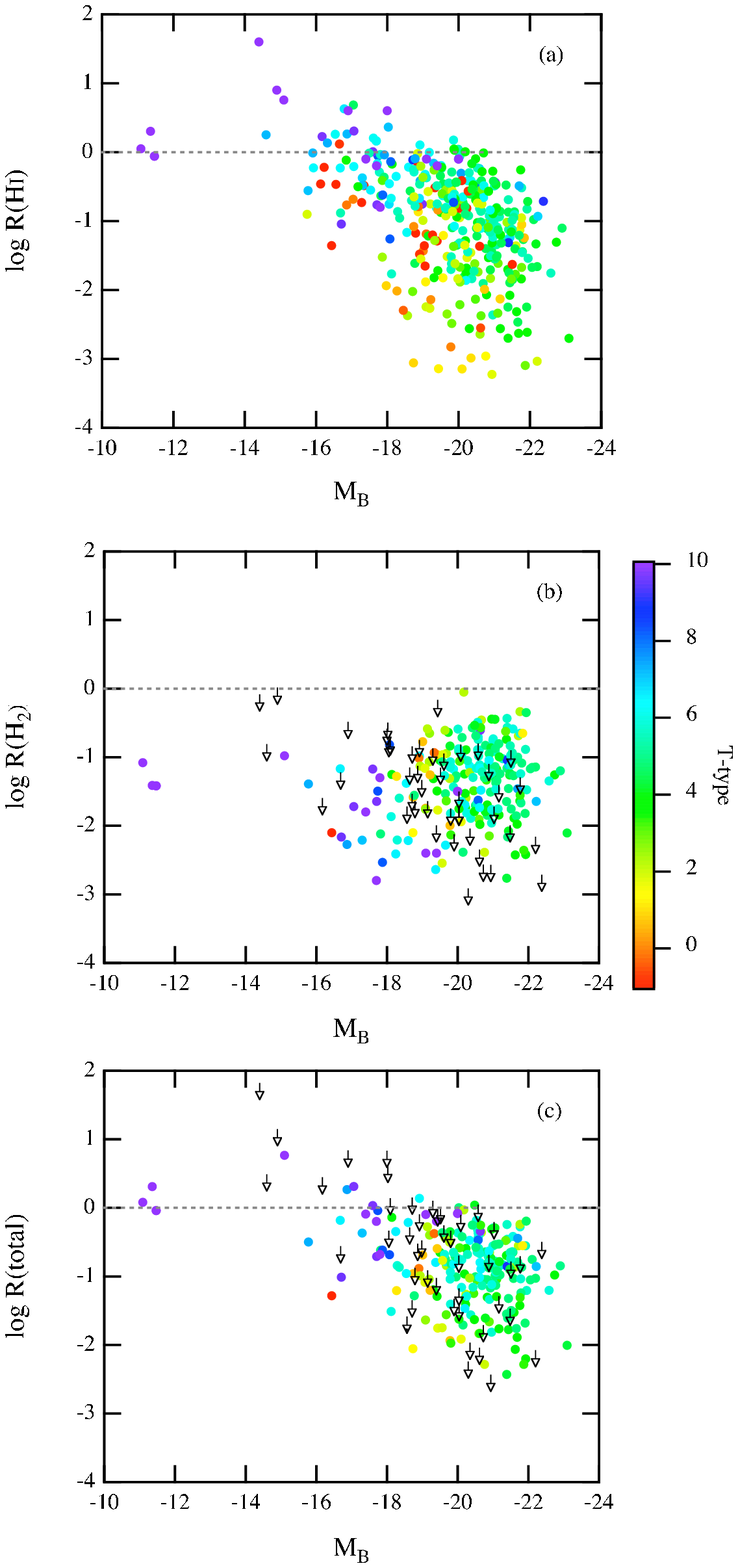}}
 \caption{Upper panel: \textit{R}(\HI) against M$_{\mathrm{B}}$  for all 392 galaxies with recorded \HI\ and H$_2$ data. Middle panel: \textit{R}(H$_2$) plotted against M$_{\mathrm{B}}$ for the same sample. Lower panel: \textit{R}(total), (defined in text) plotted against M$_{\mathrm{B}}$ for the same sample. A dashed horizontal line has been drawn at a gas mass ratio of 50\%, for ease of reference. Galaxies with only upper limits on their CO mass have been marked with arrows in panels b and c.}
 \label{fig:gas_comp}
\end{figure}

The cool, neutral phase of the ISM, composed of atomic hydrogen, acts both as a reservoir of fuel for future star formation, and as a \textit{tracer of} star formation (as new massive stars photodissociate the surrounding molecular ISM into \HI\ - e.g. \citealt{1991ApJ...366..464A}). Stars form inside molecular clouds, however, which are comprised primarily of molecular hydrogen - H$_2$ - and helium, and thus an understanding of both the neutral and molecular components of the ISM is integral to a full characterisation of the relationship between gas content and SFR. To account for the mass of helium, a standard correction is applied with most authors taking the helium mass to be a fixed fraction of the total (\HI\ + H$_2$) hydrogen mass, i.e.

\[
\mathrm{M}_{\mathrm{gas}} = \frac{(\mathrm{M}_{\mathrm{HI}} + \mathrm{M}_{\mathrm{H}_2})}{0.74}
\]

Figure \ref{fig:gas_comp} shows the relationship between \textit{R}(gas)  and B-band magnitude for the 392 galaxies with both \HI\ and CO-derived H$_2$ data. The upper, middle, and lower panels show the \HI\ mass ratio, H$_2$ mass ratio, and total mass ratio respectively, all plotted against $\mathrm{M}_{\mathrm{B}}$. Galaxies with upper limits on their CO fluxes (10\% of our sample, as described in \S2.4) have been marked with arrows in panels (b) and (c). 

Plot (a) demonstrates very similar behaviour to the larger \HI\ sample shown in Figure \ref{fig:hi} (This new set is highly incomplete at the less massive end, with only 5 galaxies with M$_{\mathrm{B}} \simgt -15$ having recorded H$_2$ masses). Late morphological types have, on the whole, larger gas mass ratios than their early-type kin, making up the purple and blue points on the upper side of the distribution. As with the larger sample, at the transition luminosity M$_{\mathrm{B}} = -19$ the distribution does show a population of gas-poor outliers, with some galaxies brighter than this having a marked decrease in their neutral gas mass ratios, down to extremely low values of $< 0.5 \%$ . 

Plot (b) shows \textit{R}(H$_2$)  for the same sample. The distribution in H$_2$ is markedly different to that in \HI: the plot shows no significant trend in \textit{R}(H$_2$) with M$_{\mathrm{B}}$, supporting the well known idea that the amount of molecular hydrogen in the ISM of low mass galaxies is much lower than their more massive counterparts (or equally that it becomes harder to detect due to variations in the H$_2$-CO relationship). There also is an apparent limit, at a molecular gas mass ratio of $\sim 50\% \; (\log \;\mathrm{\textit{R}}(\mathrm{H}_2) = 0)$, which exists independent of the mass of the host galaxy.

Plot (c) shows \textit{R}(total) for the same 392 galaxies. In less massive galaxies, plot (c) is broadly similar to plot (a), albeit with (very slightly) larger values of \textit{R}(gas). These late type galaxies are dominated by neutral hydrogen, and do not have a significant molecular component to their ISM. At the upper end, however, the molecular contribution becomes significant. The outliers, existing towards lower gas mass ratios seen in the \HI\ plots, are much less apparent, in favour of a smoother, more monotonic decrease.

\subsection{Comparison of Gas and SFR Demographics}
\label{SFR_analysis}

\begin{figure}
 \centerline{\includegraphics[scale=0.65]{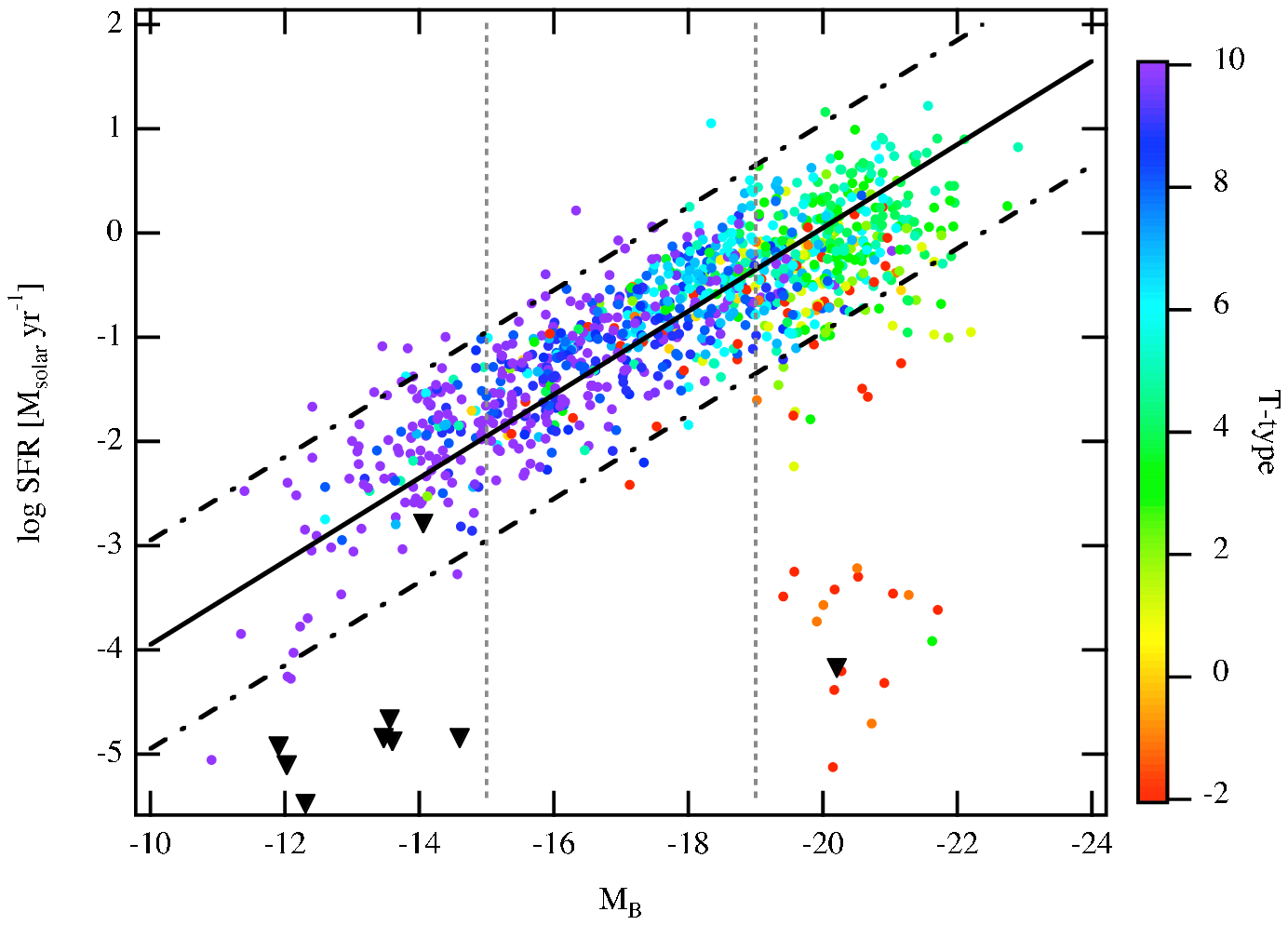}}
 \caption{Star Formation Rate plotted against absolute B-band magnitude for the complete H$\alpha$ sample. The solid diagonal line represents a value of $P* = 0$, while the upper and lower dashed lines show $P* = 1$ and $P* = -1$ respectively. The gray vertical lines at M$_{\mathrm{B}}$ = -15 and M$_{\mathrm{B}}$ = -19 indicate the boundaries separating the three SSFR behavioural regions identified by Lee et al. (2007). Galaxies with only upper limits on their H$\alpha$ fluxes are marked with downward arrows.}
 \label{fig:sfrmb}
\end{figure}

In this analysis of the SFR demographics, the sample of H$\alpha$ detected galaxies has been divided into two parts. Firstly, the entire H$\alpha$ sample - comprising 1110 galaxies - has been analysed, in order to  investigate trends as completely as possible. Secondly, in order to directly compare the SFR properties with the gas content, the sample has been restricted to only those galaxies with both H$\alpha$ and \HI\ data. This composite sample comprises 1049 galaxies (see Figure \ref{fig:venn} for clarification).

Figure \ref{fig:sfrmb} shows the SFR plotted against absolute B-band magnitude for all galaxies with available H$\alpha$ data. The vertical dashed lines represent the two transitional boundaries previously identified. The diagonal lines represent different values of the parameter $P*$, defined by \cite{2007AstL...33..283K} as 

\[
P* = \log\left(\frac{\mathrm{SFR}\cdot \mathrm{T}_0}{\mathrm{L}_{\mathrm{B}}}\right)
\] 
Where $T_0$ is the Hubble time, taken here to be $13.7 \times 10^9$ years \citep{2007ApJS..170..377S}. This parameter is analogous to the Scalo `b' parameter \citep{1986FCPh...11....1S}, in that a galaxy with $P* = 0$ is forming stars at such a rate that if it were to form stars at the current rate over one Hubble time, the total produced luminosity would equal the current galactic luminosity. By extension, values of $P* = -1$ and $P* = 1$ imply that it would take times of 10 T$_0$ and 0.1 T$_0$ respectively to fulfil the same condition.

The central solid line in Figure \ref{fig:sfrmb} represents a value of $P* = 0$, whereas the upper and lower (dot-dashed) lines represent $P* = 1$ and $P* = -1$ respectively. With the vast majority of galaxies lying between these upper and lower limits, it is possible to define a `main sequence' of star forming galaxies this way, with ($1 > P* > -1$). As expected, the mean value of $P*$ decreases with increasing luminosity. Conspicuously, however, there are many points existing as extreme outliers, with SFRs far below this `main sequence'. These occur almost exclusively in the upper (M$_{\mathrm{B}} < -19$) and lower (M$_{\mathrm{B}} > -15$) luminosity categories, with very few galaxies in the intermediate bin having aberrant SFRs. The three previously identified modes of SF behaviour are evident in Figure \ref{fig:sfrmb}: A high mass population containing some galaxies with quenched SF, which belong to the `red sequence'; a secularly evolving intermediate population, with highly regulated SFRs; and a low mass, low luminosity population, again containing members with anomalously low SFRs. Within the `main sequence' (i.e. excluding those aberrant galaxies in the extreme luminosity bins), $P*$ is lognormally distributed with a spread of $\sigma = 0.4$ dex. The mean value of $P*$ for these galaxies is $\sim-0.5$, indicating that at the current rate of SF, it would take 3 Hubble times to form a quantity of stars sufficient to produce the current luminosity (equivalent to a Scalo `b' parameter of $\sim 0.3$). 


\subsubsection{Gas and SSFR in the Local Volume}
\label{sec:LV}

\begin{figure*}
 \centerline{\includegraphics[scale=0.65]{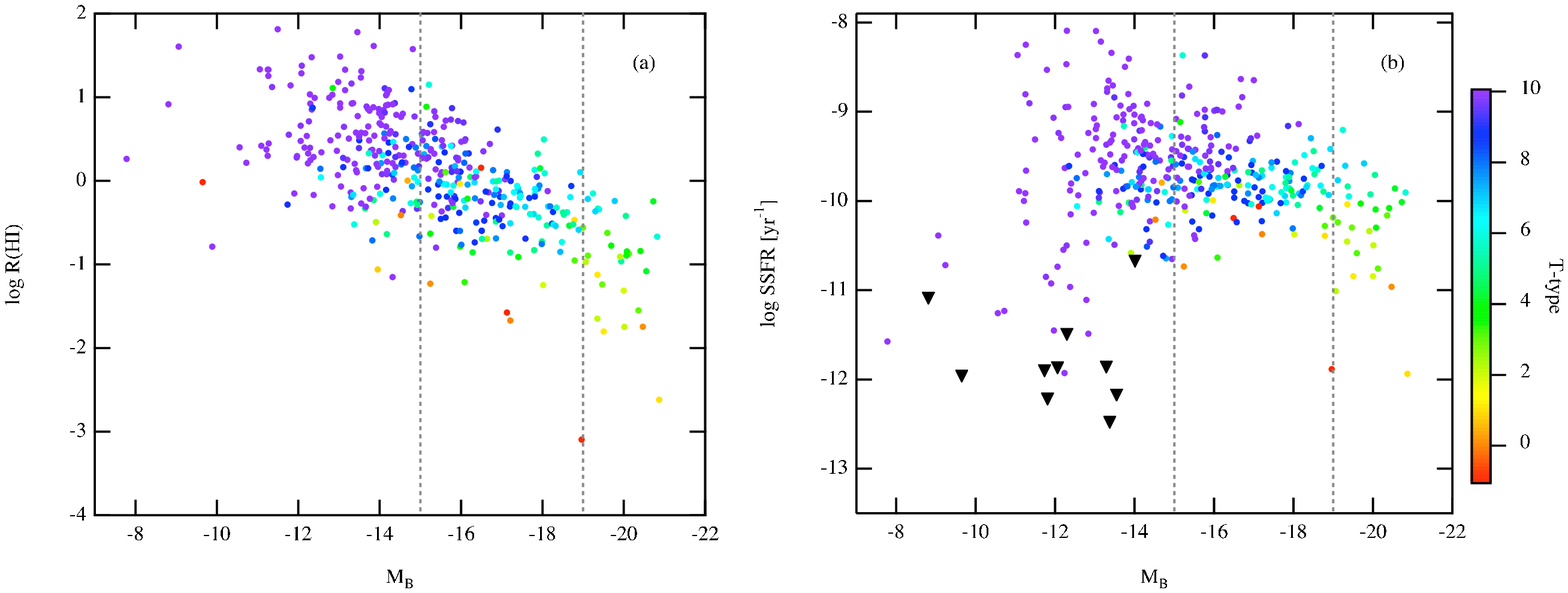}}
 \caption{Plot (a) - the relationship between \textit{R}(\HI) and the B-band luminosity for the  Local Volume galaxies described in \S1 above. Plot (b) - the relationship between the Specific Star Formation Rate and the B-band luminosity for the same sample. Dashed vertical lines at M$_{\mathrm{B}}$ = -15 and M$_{\mathrm{B}}$ = -19 indicate the boundaries separating the three SSFR behavioural regions identified by Lee et al. (2007).}
 \label{fig:kara1}
\end{figure*}



In order to be able to \textit{directly} compare distributions in SFR and gas behaviour, the sample of galaxies must be the same in each case. Figure \ref{fig:kara1}(a) shows \textit{R}(\HI)  plotted against M$_\mathrm{B}$ for galaxies in the Local Volume. 420 galaxies had a full complement of data available for analysis. Figure \ref{fig:kara1}(b) plots SSFR against M$_\mathrm{B}$ in the same way. RC3 morphological types are colour-coded as indicated in the accompanying colour bar. The dashed vertical lines at M$_\mathrm{B}$ = -15 and M$_\mathrm{B}$ = -19 indicate the boundaries separating the different behavioural regions, identified by \cite{2007ApJ...671L.113L}. The galaxies in this analysis with only upper limits on their H$\alpha$ fluxes appear as downward-facing arrows.

The distribution in \textit{R}(\HI) shown in \ref{fig:kara1}(a) shows the high luminosity outliers at low values of M$_{ \mathrm{HI}}$/M$_{ \mathrm{*}}$, mirroring the same behaviour in SSFR. It does, however, only shows a very modest indication of broadening in the region M$_\mathrm{B} > -15$.

The characteristic transitions identified in the \{EW, M$_\mathrm{B}$\}  plane by \cite{2007ApJ...671L.113L} appear in the \{SSFR, M$_\mathrm{B}$\} distribution in \ref{fig:kara1}(b). In particular, the distribution in SSFR for the least luminous (M$_\mathrm{B} > -15$) galaxies is significantly broader than for intermediate, ($-15 > \mathrm{M}_\mathrm{B} > -19$) types. At M$_\mathrm{B} < -19$, the distribution `turns over', and spreads to include lower SSFR values.


\subsubsection{Gas content and SSFR of the entire sample}
\label{sec:ES}

\begin{figure*}
 \centerline{\includegraphics[scale=0.65]{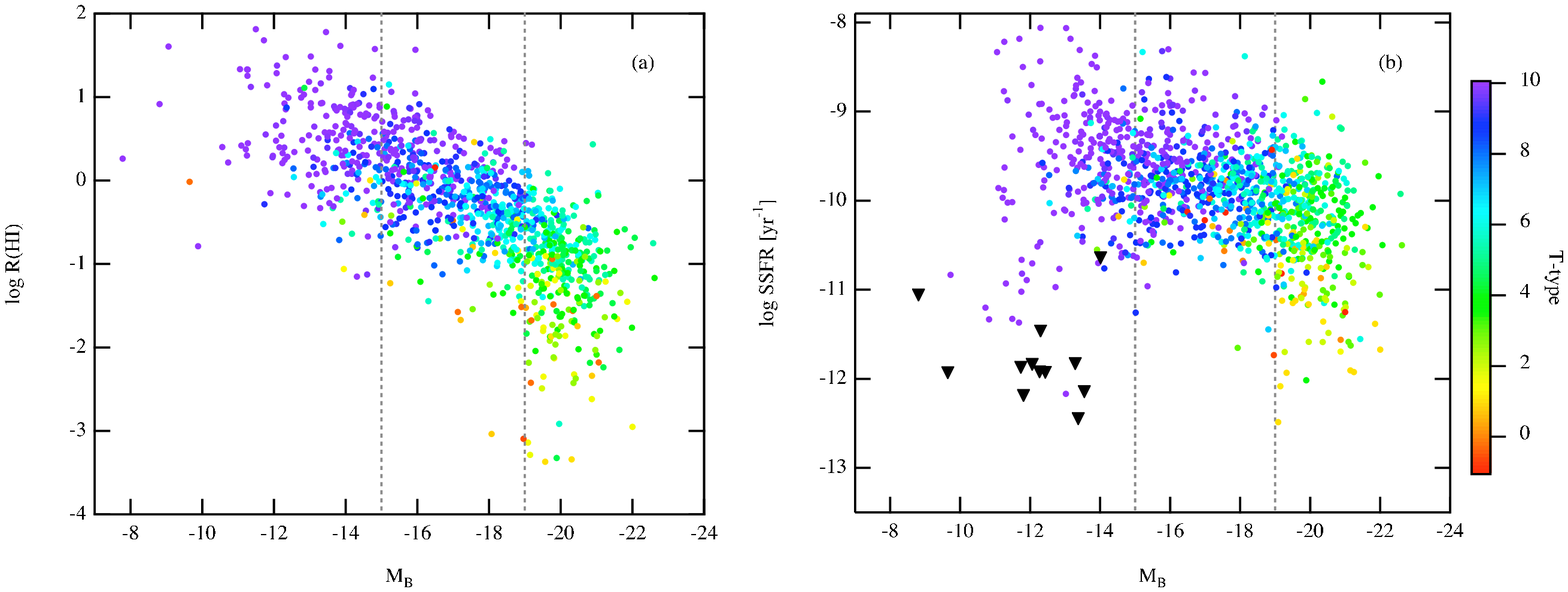}}
 \caption{Plot (a) - the relationship between the \textit{R}(\HI) and the B-band luminosity for the complete H$\alpha$ sample described in \S1 above. Plot (b) - the relationship between the Specific Star Formation Rate and the B-band luminosity for the same sample. Dashed vertical lines at M$_{\mathrm{B}}$ = -15 and M$_{\mathrm{B}}$ = -19 indicate the boundaries separating the three SSFR behavioural regions identified by Lee et al. (2007). Least-squares best fitting power laws fitted to the distributions in (a) and (b) have indices $(-0.59 \pm 0.02)$ and $(-0.25 \pm 0.03)$ respectively.}
 \label{fig:main1}
\end{figure*}

The plots shown in \S\ref{sec:LV} above have the advantage of being a very well characterised sample, with a high level of completeness for the volume studied. However, this benefit has associated weaknesses. The total volume probed by an 11 Mpc survey is very small; approximately 5000 Mpc$^3$. As such, the sample will suffer from a paucity of rarer objects, such as bright, massive galaxies, and will not be a representative sample of star formation in the $z \sim 0$ universe. H$\alpha$ surveys which probe greater volumes are required  in order to provide better coverage of the upper end of the luminosity function. To this end, the larger samples described in \S\ref{sec:samples} have been included. As discussed above, surveys specifically targeting extreme or unusual systems were excluded, in order to provide - as much as possible - a statistically representative sample of star forming galaxies at $z = 0$.

Plot \ref{fig:main1}(a) plots the \HI\ gas mass ratio against absolute B-band magnitude for this larger composite sample. M$_{\mathrm{B}}$, as opposed to the stellar mass, has been used for the abscissa, in order to facilitate easier comparison with the results of \cite{2007ApJ...671L.113L}. As in the 11Mpc sample, the distribution exhibits two characteristic transitions. The `red sequence' transition is apparent, with the galaxies in the most luminous bin (M$_{\mathrm{B}}$ $< -19$) turning off towards very low gas fractions of $< 1\%$, and becoming dominated by early types. This transition luminosity corresponds to the stellar mass transition ($3 \times 10^{10} \;\mathrm{M}_{\sun}$) discussed in \S1 above. At the less-luminous end, however, the distribution again deviates from the equivalent SSFR plot. Late type galaxies dominate at M$_{\mathrm{B}} > -17$, having large gas fractions, typically $> 50 \%$. The distribution shows no sign of the broadening towards low values displayed by the SSFR plot, as quantified below.

Figure \ref{fig:main1}(b) shows the distribution of SSFR with M$_\mathrm{B}$ for this total compilation of  H$\alpha$ data.  The distributions shown in Figure \ref{fig:main1}(a) and Figure \ref{fig:main1}(b) are consistent with those shown in  Figure \ref{fig:kara1}(a) and  \ref{fig:kara1}(b). As a qualitative check, it is also possible to make the above plots with rotational velocity instead of M$_{\mathrm{B}}$ as the x-axis mass tracer (as in \cite{2007ApJ...671L.113L}). These are not shown for the sake of brevity, but the global relationships do not differ significantly from the plots shown. 

Plot \ref{fig:main1}(b) again demonstrates the three behavioural transitions identified previously by \cite{2007ApJ...671L.113L}. Above M$_\mathrm{B} = -19$, where the population is dominated by early-type spirals, the distribution of galaxies both broadens and turns over to lower SSFR values. In the intermediate luminosity bin, $-15 > \mathrm{M}_\mathrm{B} > -19$, the population is comprised primarily of later type spirals and dwarfs. The mean SSFR is close to constant in this luminosity bin, at $\log$ (SSFR) $\sim -10.5$ yr$^{-1}$. In the lowest luminosity bin, the population is comprised almost entirely of late-type irregulars and dwarfs. The distribution of SSFRs broadens significantly, occupying a range of over three orders of magnitude ($-8.9 > \log(\mathrm{SSFR}) > -12.4$).

\begin{figure}
 \centerline{\includegraphics[scale=0.65]{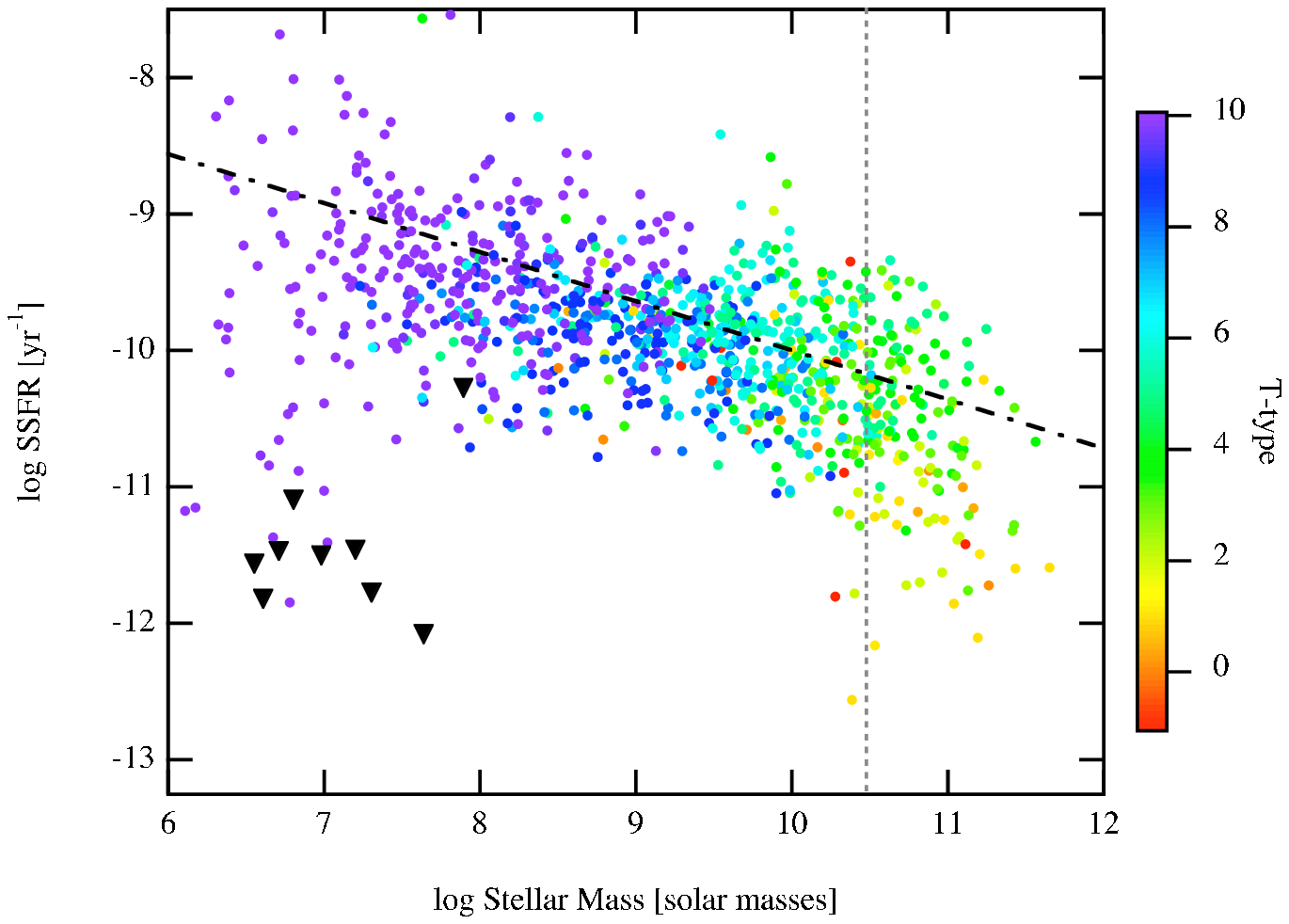}}
 \caption{The relationship between the Specific Star Formation Rate and stellar mass for the full compilation. The dashed vertical line indicates the transition mass (identified previously at $\sim3\times 10^{10}$M$_{\sun}$). The dashed sloped line represents the 'SF Sequence' defined by Schiminovich et al. (2007)}
 \label{fig:stellarmass}
\end{figure}

Figure \ref{fig:stellarmass} shows, for reference, the changing distribution of SSFR in the stellar mass plane (as opposed to using M$_{\mathrm{B}}$ as the abscissa). The distribution with stellar mass is very similar to that in the SSFR-M$_{\mathrm{B}}$ plane, including both the upper and lower transitions, and the interim behaviour. The `upper' transition is seen to manifest very close to the transition mass, $\sim3\times10^{10}$M$_{\sun}$ previously identified (e.g. \citealt{2007tS..173..267S}; \citealt{2007ApJS..173..315S}), and for reference a vertical line has been drawn at this mass. The line defining the `star forming sequence', defined by \cite{2007ApJS..173..315S} as
\[
\log \;\mathrm{SFR/M}_* = -0.36 \log \mathrm{M}_* - 6.4
\]
has been overplotted. Figure \ref{fig:stellarmass} is consistent with the high mass turnover results found recently (e.g. \citealt{2007tS..173..267S}), which are generally found using larger samples biased towards more massive galaxies.

\begin{figure}
 \centerline{\includegraphics[scale=0.42]{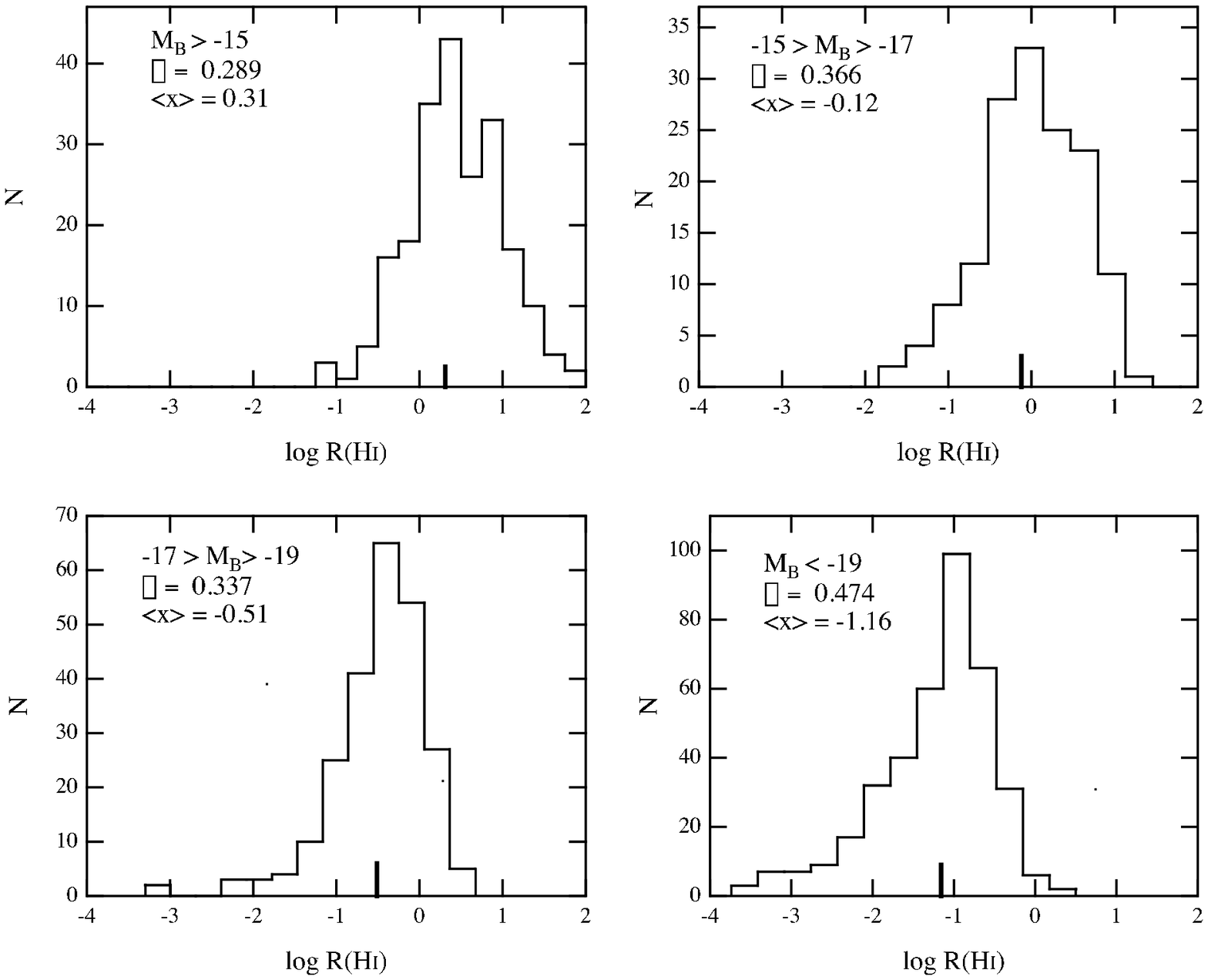}}
 \caption{Histogram showing the distribution of \textit{R}(\HI) for the complete H$\alpha$ sample shown in Figure \ref{fig:main1}(a) above. The sample has been divided into four luminosity bins as shown in the diagram. The standard deviation ($\sigma$) and the mean ($<$x$>$) are also shown; the position of the mean is marked on the plot with a black line. The value of $\sigma$ has been corrected to account for the slope of the distribution as detailed in the text.}
 \label{fig:hist2}
\end{figure}

\begin{figure}
 \centerline{\includegraphics[scale=0.42]{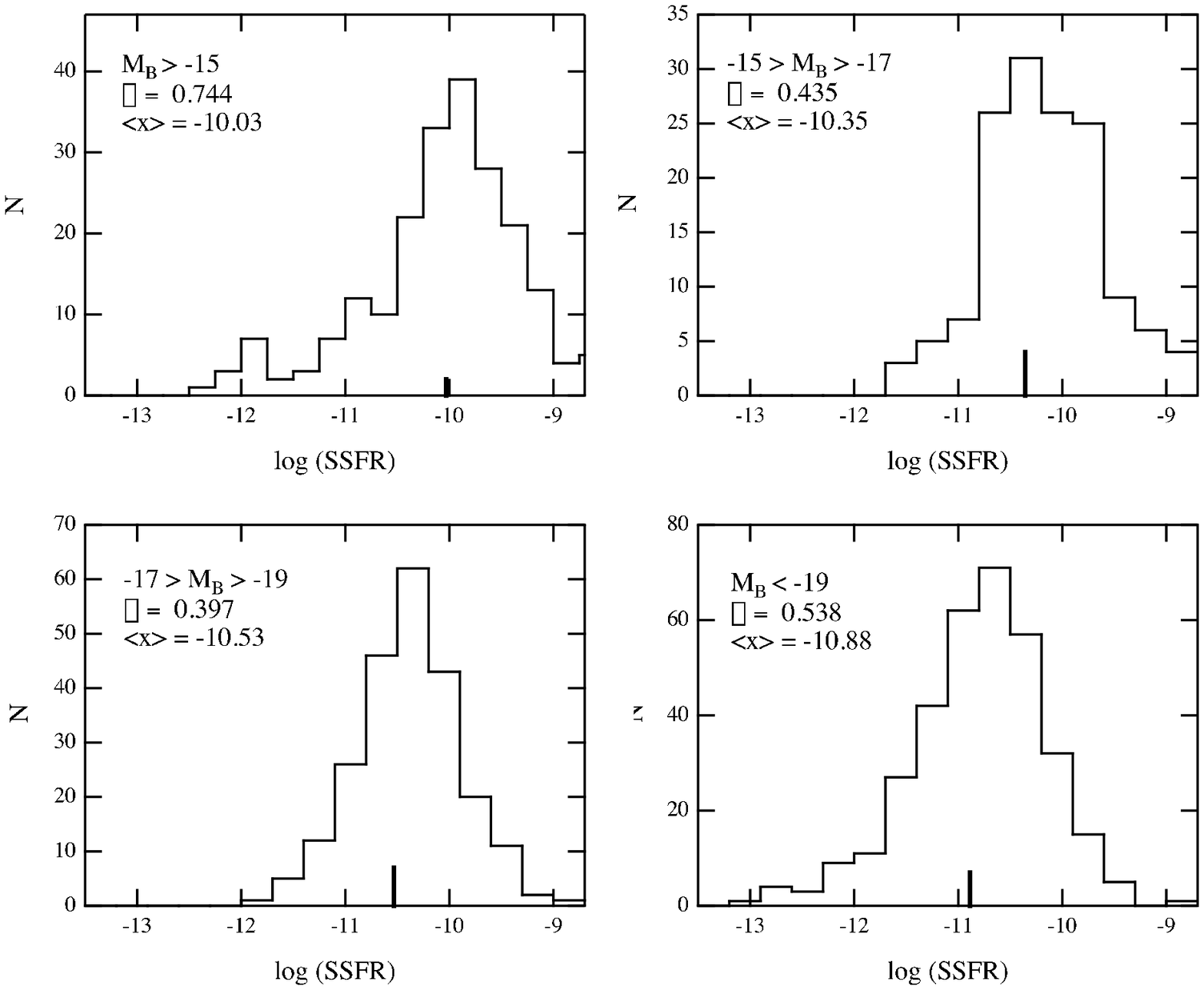}}
 \caption{Histogram showing the distribution of SSFR for the complete H$\alpha$ sample shown in Figure \ref{fig:main1}(b) above. The sample has been divided into four luminosity bins as shown in the diagram. The standard deviation ($\sigma$) and the mean ($<$x$>$) are also shown; the position of the mean is marked on the plot with a black line. The value of $\sigma$ has been corrected to account for the slope of the distribution as detailed in the text.}
 \label{fig:hist1}
\end{figure}

The changing distributions of SSFR and \HI\ gas mass ratios are shown in more detail in Figures \ref{fig:hist2} and \ref{fig:hist1}. The data are divided into four luminosity bins, taking the upper and lower marked regimes as two separate bins, and subdividing the middle ($-15 > \mathrm{M}_\mathrm{B} > -19$) bin into two equal sections. In Figure \ref{fig:hist2}, the distributions of \HI\ gas mass ratio in the four bins are shown as histograms. In Figure \ref{fig:hist2}, the equivalent histogram is shown for the SSFR distribution. The legends display the standard deviation of the data in the respective bins. 

Figure \ref{fig:hist2} displays the high-luminosity broadening in \textit{R}(\HI), and the lower values of $\sigma$ in the intermediate bins, mirroring similar behaviours in the respective bins of Figure \ref{fig:hist1}. There is significant difference, however, in the lowest luminosity bin. Whereas the distribution of SSFRs in this region becomes much broader than any other region, the distribution of \HI\ gas mass ratios is \textit{tighter} in this region than elsewhere. The distributions in \HI\ gas mass ratio are also negatively skewed, as opposed to the more symmetrical distributions in SSFR. 

Figure \ref{fig:hist1} shows the SSFR distribution in the most luminous class to be broader than the two intermediate classes, corresponding to the already-noted 'turnover' towards low SSFRs. The intermediate bins are lognormally distributed, with lower values of $\sigma$ than either extreme bin. The lowest luminosity bin, containing the dwarf galaxies with anomalous SF behaviour, has the highest of all standard deviations, and a highly non-lognormal distribution. 


Another significant difference between Figures \ref{fig:hist1} and \ref{fig:hist2} is the evolution of the mean. Both the mean SSFR and the mean \HI\ mass ratio decrease with luminosity, but by significantly different amounts. The difference in mean SSFR between the two most extreme bins is 0.72 dex; this is contrasted with a 1.37 dex difference in mean \textit{R}(\HI) over the same range. This can be more explicitly seen in the difference between slopes for the best fitting power law: the \textit{R}(\HI) distribution in Figure \ref{fig:main1}(a) has a power law fit of index $N = -0.59 \pm 0.02$ (essentially identical to that in Figure \ref{fig:hi}), whereas the distribution in SSFR is significantly shallower, at $N = -0.25 \pm 0.03$.  It is worth noting in Figures \ref{fig:hist2} and \ref{fig:hist1} that for all bins save the least luminous, the value of $\sigma$(\HI) is systematically greater than $\sigma$(SSFR). This likely results from the slopes of the distributions artificially introducing scatter across the relatively wide (2 magnitudes, 0.8 dex in luminosity) bins. This effect can be removed by subtracting 0.235 from $\sigma$(\HI) and 0.1 from $\sigma$(SSFR), which removes the mean effect of the slopes discussed below. When this is taken into account, $\sigma$(\HI) is lower than $\sigma$(SSFR) in all bins. 

\begin{figure}
 \centerline{\includegraphics[scale=0.7]{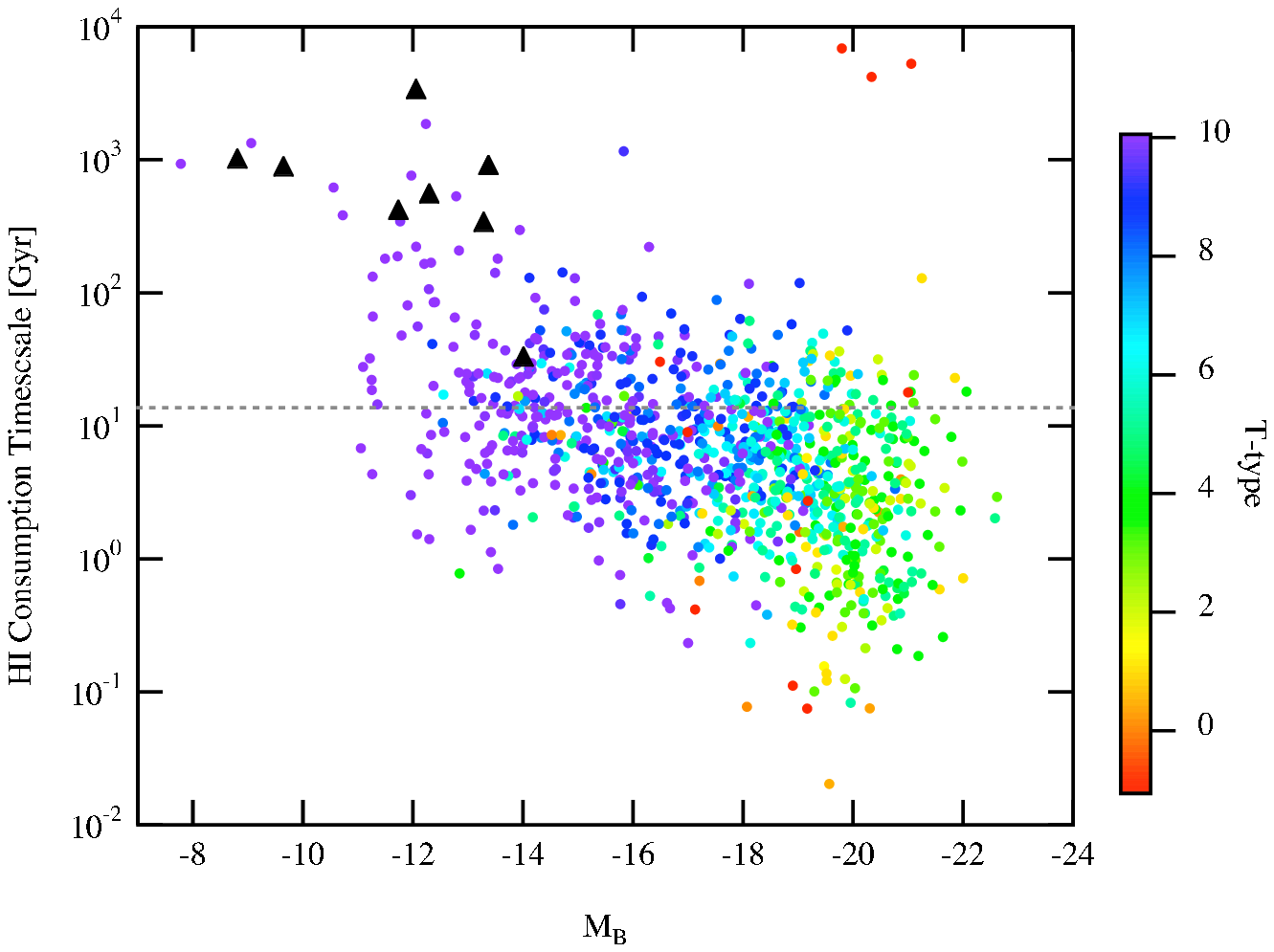}}
 \caption{Characteristic \HI\ consumption timescale (defined as M(\HI) / SFR and expressed in Gyr). The dashed horizontal line represents the Hubble time. These timescales were derived using a Salpeter IMF - timescales should be increased by 1.44 for a Kroupa IMF \citep{1993MNRAS.262..545K}}
 \label{fig:consume}
\end{figure}

This variation in slope can be seen as a change in the mean \HI\ consumption timescale (as discussed by, for example, \citealt{1969AJ.....74..859R}; \citealt{1986A&A...161...89S}; \citealt{1983ApJ...272...54K}): with increasing luminosity, the \HI\ content drops off faster than the SFR, resulting in shorter \HI\ consumption timescales  for more luminous galaxies. This is shown explicitly in Figure \ref{fig:consume}, where the mean \HI\ consumption time decreases monotonically with increasing luminosity (upward facing arrows denote galaxies with only upper limits on their SFRs). An interesting feature of Figure \ref{fig:consume} is the lower limit - with the exception of one galaxy (NGC 2681), all points have \HI\ consumption times $>$100 Myr, which is (approximately) one dynamical time for a typical galaxy. This lower cut-off effect is caused by the SFR ceiling, found to be roughly equal to the gas mass divided by the minimum time to assemble the gas - i.e. the free-fall collapse time. Naturally, therefore, there should be a rough lower limit to the consumption time equal to the dynamical time.

The \HI\ consumption timescales present in some low luminosity dwarf galaxies are in some cases extremely high - for example, the dwarf galaxy UGC 07298 (M$_{\mathrm{B}}$ = -12.27) has a \HI\ consumption time of $2.7 \times 10^{12}$ years. These long consumption times likely result from a combination of a very low SFR and an extended \HI\ disc, where a large fraction of the \HI\ disc, though recorded by the single dish measurements, is not actively available for star formation. These two factors can conspire to produce an extremely long consumption time.

\subsubsection{Correlation of SFR with different ISM components}
 It is a matter of ongoing debate whether the SFR correlates better with H\,\textsc{i} or H$_2$ density; \cite{1994A&A...292....1B}, \cite{1994A&A...289..715D}, and \cite{1998ApJ...498..541K} find a significantly better correlation with the atomic gas density. \cite{1998ApJ...498..541K} suggest that the large proportion of the ISM composed of H\,\textsc{i} is responsible for the better correlation with \HI\, while variations in the CO / H$_2$ conversion are responsible for the poor correlation with molecular gas. 
However, \cite{2002ApJ...569..157W}  investigated the correlation between SFR density and gas density and found that $\Sigma_{\mathrm{SFR}}$ correlated much better with $\Sigma_{\mathrm{H_2}}$ than with $\Sigma_{\mathrm{H\,\textsc{i}}}$, to the extent that their H\,\textsc{i} data were inconsistent with a Schmidt law, while their CO data showed excellent correlation. (They suggest that this is due to H\,\textsc{i} self shielding and converting to H$_2$ at higher column densities, as described by, for example, \citealt{1979ApJ...227..466F}; \citealt{1987ApJ...319...76S}.) \footnote{It is often the case that global studies find a superior connection with \HI\, while resolved studies of individual galaxies find H$_2$ to correlate better: see \cite{2009arXiv0903.3950F} for a recent discussion of this interesting problem.}

In order to more rigourously quantify the relation between gas content and SSFR, a correlation function will be used: being the standard test of correlation, the Pearson product-moment correlation coefficient (PPMCC) will be used\footnote{Is is also possible to use a non-parametric statistic, such as Spearman's Rank - both measures of correlation give virtually identical results.}. This is defined as

\[
\mathrm{r}_{\mathrm{xy}} = \frac{n\sum x_i y_i - \sum x_i \sum y_i}{\sqrt{n\sum x_i^2 - (\sum x_i)^2}\sqrt{n\sum y_i^2 - (\sum y_i)^2}}
\]

\[
\:\:\:\:\:\;\:= \frac{1}{n-1} \;\;\sum_{i = 1}^n \; \left(\frac{x_i - \bar{x}}{\sigma_x}\right)\left(\frac{y_i - \bar{y}}{\sigma_y}\right)
\]
where $r_{xy}$ is the correlation coefficient, $\sigma$ is the standard deviation, and $n$ is the number of points. The three dimensional parameter space (= \textit{R}(gas), SSFR, M$_{\mathrm{B}}$) has been sliced into 1 mag bins along the M$_{\mathrm{B}}$ `axis', and the resulting correlation between \textit{R}(gas) and SSFR calculated for members of each bin separately. This way, we explicitly calculate the variation of the dependence of SSFR on \textit{R}(gas), as M$_{\mathrm{B}}$ changes.

\begin{figure}
 \centerline{\includegraphics[scale=0.43]{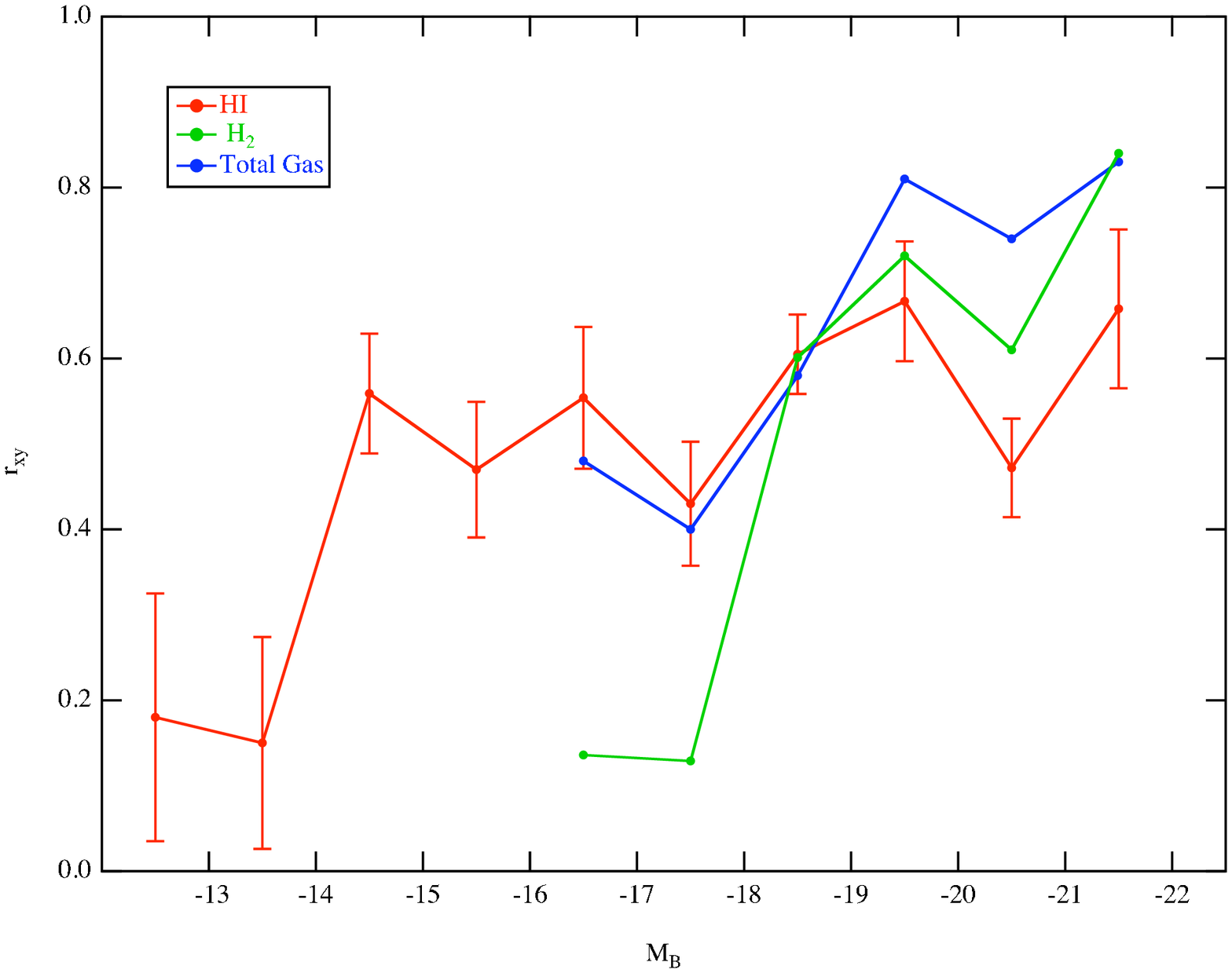}}
 \caption{The variation of the Pearson product-moment correlation coefficient with absolute B-band magnitude, for correlations between different gas mass ratios (\HI, H$_2$ traced via CO, and total gas) and SSFR. Error bars were estimated using a Monte Carlo bootstrapping method. Due to the low numbers of low-luminosity galaxies, galaxies with M$_{\mathrm{B}} > -13$ were considered as one group, as were galaxies with M$_{\mathrm{B}} < -21$.}
 \label{fig:corrgraph}
\end{figure}

Due to the low numbers available at the extreme ends, galaxies with M$_{\mathrm{B}} > -13$ were considered to be a single group, as were galaxies with M$_{\mathrm{B}} < -21$. Figure \ref{fig:corrgraph} shows the variation of the PPMCC (between SSFR and gas mass ratio) with absolute magnitude, for the three definitions of gas ratio discussed above. The red line shows the correlation between \textit{R}\HI\ and SSFR - in order to include as much data as possible, the sample drawn upon for this calculation was the full sample, shown in Figure \ref{fig:main1}. The green and blue lines show the value for \textit{R}(H$_2$), and \textit{R}(total) respectively. Due to the lack of CO data at the lower luminosities (see Figure \ref{fig:gas_comp}), the plot above has been truncated below M$_{\mathrm{B}} = -16$ for the molecular and total gas data. 1$\sigma$ errors - or more precisely robustness measures, being as they were estimated using a Monte Carlo bootstrapping method - were typically $< \pm 0.1$, and are shown for the \HI\ data. For datasets smaller than $\sim 20$ points, the bootstrap method of error estimation becomes unreliable: for this reason, bootstrap error bars were not calculated for molecular and total correlations.



\section{Discussion}
\subsection{Error Budget}
\label{sec:err}

Each observational parameter will come with some associated error, as will each variable conversion. As a result, it is imperative to assess how much scatter in the relationships will be driven by these errors, and how much results from innate parametrical variance. 

Photometric errors are, in general, small, being typically $\pm$ 0.1 mag (with the exception of the faintest dwarf galaxies, which can have photometric errors of $\sim$0.3 mag), as are errors on H$\alpha$ and \HI\ fluxes. H$\alpha$ extinction correction errors are also, in general, small, though the scatter in the statistical relationship does increase as a function of luminosity. While the scatter for the least luminous dwarf galaxies (which have very small levels of extinction) is negligable ($\sim$8\%), the scatter at the luminous end (M$_{\mathrm{B}} < -19$) is $\sim$40\%. The scatter between these extremes is $\sim$20\%.

By far the dominant source of error in this analysis is the adoption of an assumed colour in order to calculate the stellar mass. There is a $\sigma$ of $\sim 0.3$ dex in the relationship shown in Figure \ref{fig:mass}, which far outweighs the other error sources. As noted previously, however, the demographic results of the analysis are identical when only galaxies with true colours are considered; the colour adoption method will generally cause scatter to \textit{reduce}, and indeed the most extreme points in all the plots in this analysis are derived from galaxies with true colours. We can say, therefore, that the predominant cause of the scatter in the diagrams is true cosmic scatter, and an estimate of the error in gas and SSFR parameters would be $\pm 0.3$ dex.

\subsection{Gas Distribution}
Figure \ref{fig:hi} shows the distribution of the normalised \HI\ content for a very large sample of low redshift galaxies, across the entire luminosity function. It reproduces the long known result (e.g. \citealt{1994ARA&A..32..115R}) that earlier type galaxies, at the more luminous end of the spectrum, have far less gas per unit stellar mass than their less massive companions, having long ago converted the ISM material into stars. The least massive galaxies, on the other hand, have yet to exhaust - or indeed fully tap - the \HI\ available in the ISM. This is, of course, in accordance with the downsizing paradigm, whereby massive galaxies finish their star forming era at higher redshifts, whereas the mean star formation rate of dwarf galaxies has changed little over cosmic time. 

This investigation is somewhat hampered by the paucity of massive very early type galaxies, which comprise the classic red sequence. It is possible to discern, however, a small population of massive galaxies (M$_{\mathrm{B}} > -19$), which `turn over' towards very low \HI\ mass ratios of $< 1\%$. This suggests that the red-blue sequence divide is at least connected to the availability of \HI\ in the ISM.

The three plots in Figure \ref{fig:gas_comp} show the variation in \textit{R}(gas) with the species of gas being considered. Panel (a) confirms that the \HI\ properties of this restricted sample (only galaxies with both \HI\ and H$_2$ data are included here) are consistent with the larger sample shown in Figure \ref{fig:hi}. Panel (b) shows the equivalent plot for the normalised molecular gas content. Immediately obvious is the fact that there is no strong trend in \textit{R}(H$_2$) with M$_{\mathrm{B}}$, as there is in \HI; the dwarf galaxies have normalised gas contents similarly low as the more massive spirals. This is, of course, an affirmation of the result that dwarf galaxies are \HI\ dominated, with the molecular contribution to their ISM being small, and has been noted previously (e.g. \citealt{1991ARA&A..29..581Y}). It does, however, rely on the assumption of the infallibility of the CO-H$_2$ ratio, which some authors have theorised to become unreliable in low mass (hence low metallicity) systems. Work on calibrating the conversion factor to low metallicity systems is ongoing (e.g. \citealt{2009arXiv0901.2526O}). Even with a more sophisticated calibration, however, it is unlikely that the molecular component of the lowest mass galaxies is comparable to the \HI\ content. 

More interesting is the existence of an apparent maximum H$_2$ content of approximately 30-50\% (log \textit{R}(H$_2) \sim$ 0), where no such maximum is seen in the \HI\ distribution. \HI\ and H$_2$ display very different behavioural features when spatially mapped - while the H$_2$ distribution tends to be centrally peaked, falling off with radius, \HI\ typically has a much more flat profile, often having a constant surface density out to many times the optical radius (e.g. \citealt{1982ARA&A..20..517M}; \citealt{1989ApJ...344..171K}; \citealt{1991ARA&A..29..581Y}; \citealt{2008MNRAS.383..809B}). Single dish integrated measurements of the \HI\, then, measure the total neutral hydrogen content of a galaxy, not all of which may be actively involved in the star formation process. It is possible that many galaxies (particularly dwarf galaxies which lack the ability to gravitationally collapse them) have large \HI\ reserves inhabiting their outer discs, which is unavailable for star formation: this would lead to the extremely high \HI\ contents (log \textit{R}(\HI) $> 1$) seen in Figures \ref{fig:hi} and \ref{fig:gas_comp}. By contrast, global integrated measurements of the molecular gas mass only measure the dense circumnuclear gas, which suffers from this apparent maximum. This limit may result from a gravitational threshold effect, whereby having a large reserve of H$_2$ in the ISM without concomitant star formation is simply not a stable situation for a galaxy. Alternatively, it may be that photodissociation effects are the cause of the limit, whereby the interstellar UV radiation field photodissociates any large agglomerations of molecular gas (e.g. \citealt{2004ApJ...609..667S}; \citealt{2009ApJ...699..850K}). The origin of this threshold is not fully understood, however, and a full exposition of the various theories certainly lies beyond the scope of this work.


Panel (c) shows the distribution of the normalised \textit{total} gas content, calculated by adding the \HI\ and H$_2$ components and applying a standard helium correction. Unsurprisingly, at low masses, this plot is essentially identical to the equivalent \HI\ plot, as the galaxies in question are highly \HI\ dominated. As the galaxies increase in luminosity, the \HI\ component decreases while the H$_2$ gas mass ratio shows no significant trend with M$_{\mathrm{B}}$, with the result that galaxies become more H$_2$ dominated with increasing mass. The distribution also tightens compared to the \HI\ locus in panel (a). In addition, the gas-poor outliers are much less apparent, although this may result from the uncertainty introduced by the CO upper limits - in the absence of conclusive CO masses for these galaxies, it cannot be said with certainty that the downturn is absent. 


\subsection{Star Formation Properties and Modality}
As explained in \S \ref{SFR_analysis}  above, Plot (b) of both Figures \ref{fig:kara1} and \ref{fig:main1} displays essentially the same trends as identified by \cite{2007ApJ...671L.113L}, particularly with respect to the two transitions, at M$_{\mathrm{B}} < -19$ and M$_{\mathrm{B}} > -15$. These transitions are also apparent when considering the absolute SFR (Figure \ref{fig:sfrmb}), with many galaxies in these two magnitude classes falling outside of a `main sequence' of star forming galaxies. We address these two transitions in turn. 

At the upper end of the luminosity function, the H$\alpha$ based SSFRs exhibit the characteristic turnover towards low values ($< 10^{-12}$ yr$^{-1}$), indicating that their star formation is being suppressed. 
It is, however, important to note that this turn over does \textit{not} manifest in the distributions of the total (atomic plus molecular) gas ratio - which as Figure \ref{fig:corrgraph} shows is better correlated with SSFR at the high mass end than \HI. It is likely that this paradoxical situation is caused by the fact that the number of quenched galaxies is small ($\sim 25$) compared to the number of galaxies in the bulk of the distribution. Figure \ref{fig:corrgraph} shows that for the majority of high luminosity galaxies (making up the bulk of the distribution), their SFR is being driven by the molecular component of the ISM. However, the lack of a break in the molecular (and total) gas distribution (Figure \ref{fig:gas_comp}, panels b and c) suggests that while the main sequence of star formation is driven by the molecular component, quenching events that suppress star formation primarily affect the \HI\ content, which leads to the mirrored `turnovers', seen in both SSFR and f$_{\mathrm{HI}}$.

\subsubsection{Star Formation in Dwarf Galaxies}
We move now to the likely causes of the large spread in the specific star formation rate distribution of the low luminosity dwarf galaxies. The  possible causes of this behaviour can be categorised under two main headings; firstly, the various physical mechanisms that could cause the star formation properties of dwarf galaxies to not only exhibit a large spread, but also be decoupled from their gas contents. The second possibility is that at low luminosities, the star formation rate is not being faithfully traced by the H$\alpha$ luminosity, leading to a mischaracterisation of the SFR. These will be addressed in turn.

There has been much work dedicated to the processes involved in star formation at low masses (\citealt{1973ApJ...179..427S}; \citealt{1980ApJ...242..517G}; \citealt{1986ApJ...303...39D};  \citealt{1998ApJ...493..595H}, and \citealt{1999ApJ...513..142M}, to name just a few). Much of this work has centred on the role of kinematical disturbances: being as they are small systems, dwarf galaxies are inherently more susceptible than their more massive counterparts to various types of disruption. Their shallower potential wells leave them open to disturbances resulting from both internal and external events and processes which could act to change the star formation behaviour. \cite{1973ApJ...179..427S} first noticed that dwarf galaxies exhibited a distinctly different mode of star formation compared to `normal' spirals, being dominated by a `bursting' mode, and \cite{1980ApJ...242..517G} argued that this bursting mode results from internal feedback processes; supernovae (the rate of which is proportional to SFR) in low mass galaxies have a large chance of disrupting the ISM so that SF is quenched, until it can re-collapse enough to trigger another star forming event. 

\cite{2007ApJ...667..170S} ran closed box smooth particle hydrodynamical simulations of low mass galaxies ($3.18 \times 10^8 \;\mathrm{M}_{\sun} - 8.6 \times 10^9 \;\mathrm{M}_{\sun}$), finding that the SFR in these low mass galaxies exhibited periodic fluctuations (which they termed `breathing'), and that the amplitude of these fluctuations varied inversely with the size of the galaxy. However, it is unclear whether these processes are sufficient to explain the range of behaviours displayed by the dwarf galaxies in our sample. The simulations of \cite{2007ApJ...667..170S}, for example, show only a range of $\sim$ 1 dex in SFR (ranging from $1 \times 10^{-4}\; \mathrm{M}_{\sun} \;\mathrm{yr}^{-1}$ to $2 \times 10^{-3}\; \mathrm{M}_{\sun} \; \mathrm{yr}^{-1}$) in the lowest mass galaxy considered - insufficient to explain the full $\sim$ 3 dex range in SFR (and SSFR) in Figure \ref{fig:sfrmb} and \ref{fig:main1}.

The possibility must then be considered that the true SFR in dwarf galaxies is not being reliably  measured by standard linear conversions of the H$\alpha$ luminosity. A number of possible causes have been discussed in previous work, with recent attention being focussed on stochasticity in the formation of high mass, ionizing stars, and the potential of systematic variations in the IMF (e.g., \citealt{2007ApJ...671.1550P}; \cite{Lee_09}). In terms of stochasticity, the upper end of the IMF will only be sparsely sampled in galaxies with low SFRs. If the mean time between formation of an O star is greater than the mean lifetime of an O star ($\sim$5 Myr), then there will be times where the galaxy does not host an O star, but lower mass stars are being formed. This will result in episodic H$\alpha$ emission, unrelated to the true underlying SFR, or other physical properties such as the gas or stellar mass.  Beyond pure statistical effects, there has also been recent debate on whether the IMF in dwarf and low surface brightness galaxies could be systematically deficient in high mass stars.
    
It may be that some combination of such effects are required in order to explain our results in the low mass regime. The low mass group contains many galaxies with high SSFRs, which are rapidly forming gas into stars, along with the galaxies with apparently quenched SSFRs. In Figure 5, for example, the low mass group contains both galaxies on the `main sequence' and galaxies with extremely low SFRs of $<$ $10^{-5}$M$_{\sun}$yr$^{-1}$ . In one possible scenario, the effect of  variations in the star formation histories of low mass galaxies on the H$\alpha$ luminosity are amplified by stochasticity in the formation of high mass stars. During active periods of star formation, the H$\alpha$ luminosity is more likely to accurately trace the SFR, whereas during relatively quiescent periods, when the SFR drops below a certain threshold ($\sim$10$^{-3}$ M$_{\odot}$ yr$^{-1}$),  the H$\alpha$ emission drops dramatically, creating the illusion of more quenching than actually exists. Such issues are discussed in greater detail by Lee et al. (2009) and Tremonti et al. (2009, in preparation).

\subsubsection{The ISM-SFR connection} 

Figure \ref{fig:corrgraph}, as explained in the text, shows the variation of the PPMCC with absolute magnitude for the three definitions of \textit{R}(gas). \HI\ data being available for many more galaxies, the correlation between normalised \HI\ and SFR has been shown for the entire range of luminosities - the correlation between normalised H$_2$ and SFR, and total gas content and SFR has been shown where data are available.

The correlation between \HI\ and SFR is essentially the same for all bins with M$_{\mathrm{B}} < -15$ - it then abruptly drops at the lowest magnitudes -  indicative of the extreme broadening in SSFR (discussed above), which is not apparent in the \HI\ distribution. In high mass galaxies  M$_{\mathrm{B}} < -19$, the correlation with the molecular component is found to be better than the neutral component. For galaxies with M$_{\mathrm{B}} < -19$, however, the correlation with the molecular component drops to negligible values, with the correlation with the \textit{R}(total) being determined entirely by (and therefore approximately equal to) the value for \HI. The best correlation for the most luminous bin  of galaxies (M$_{\mathrm{B}} < -19$) is found to be with the \textit{total} gas ratio - the sum of \HI\ and H$_2$, corrected for He. 



%

\section{Conclusions}
In this work, we have presented an analysis of the gas content (both molecular and neutral) and H$\alpha$ based star formation rate of a sample of galaxies representative of the overall galaxy population in the local universe. Several archival datasets and surveys were drawn upon in order to compile our sample, including the H$\alpha$ survey of the Local Volume \citep{2008ApJS..178..247K} which provides statistically complete H$\alpha$ coverage of field galaxies to masses below that of the SMC for the first time, allowing a full characterisation of the star formation behaviour in all types of local galaxies. The final sample of around a thousand local galaxies is treated in a homogenous and consistant manner, allowing star formation rates, gas contents, and stellar masses to be compared in a uniform manner over a mass range of over 5 orders of magnitude. Our main conclusions are as follows:

\begin{itemize}
\item{
\HI\ gas mass to stellar mass ratios (\textit{R}(\HI), defined as M$_{\mathrm{HI}}$/M$_*$) were analysed for a large sample of $z \sim 0$ galaxies. \textit{R}(\HI) was found to decrease with increasing luminosity, in line with previous results. A small population of galaxies at the high mass end have anomalously low \textit{R}(\HI), resulting in the distribution having a population of distinct outliers at M$_{\mathrm{B}} \sim -19$. 
}\\
\item{
The molecular gas mass ratio (\textit{R}(H$_2$) $\equiv$ M$_{\mathrm{H2}}$/M$_*$) does not display the monotonic decline with increasing luminosity exhibited by the neutral gas. Moreover, there appears to a a maximal value of the molecular gas to stellar mass ratio (which does not exist in \HI) at $\sim$30-50\% of the stellar mass.  The total gas distribution shows a decline with increasing luminosity, but no outliers; those galaxies with the lowest \HI\ masses have the largest H$_2$ corrections.
}\\
\item{
The rate of star formation traces a monotonically increasing sequence with luminosity, with some distinct exceptions. At the low mass end, many galaxies appear to have extremally low SFRs (if the standard linear H$\alpha$ luminosity - SFR relationship is to be accepted), falling off a `main sequence' of galaxies defined by $1>P*>-1$. Intermediate luminosity galaxies fall within this sequence, almost without exception. At the high luminosity end, in addition to a slow and general trend towards lower SFRs (and $P*$ values), there is a distinct population of galaxies with `quenched' SF, with extremely low SFRs.
}\\
\item{
When directly comparing gas mass ratios with SSFRs (so both SFR and M$_{\mathrm{HI}}$ normalised to the stellar mass) for identical samples of galaxies, it is clear that the population of low mass galaxies with a wide spread in SSFRs is not matched by an equivalent shift to a larger range of \textit{R}(\HI) - the dwarf galaxies with extremely low SFRs do not, apparently, suffer from a dearth of fuel for SF in their ISM. At the high mass end, the SSFR matches the \HI\ distribution well, suggesting that the lowering of the SFR in massive galaxies is intimately connected to the \HI\ content. 
}\\
\item{
Although our restricted samples preclude a full analysis of the origins of the red / blue sequence bimodality, it must be remarked that the results presented herein are consistent with the finding that the transition between the red and blue sequences is, at the very least, intimately connected with changes in the available gas supply, and may in fact be primarily driven by the amount of gas available to fuel star formation.
}\\
\end{itemize}

\section*{Acknowledgments}
We would like to thank Ben Johnson for helpful discussions, and David Spergel for the original idea behind the project.  We would also like to extend our thanks to the anonymous referee for their insightful comments, which helped improve this paper. MSB is supported by STFC. This research has made use of the NASA/IPAC Extragalactic Database (NED) which is operated by the Jet Propulsion Laboratory, California Institute of Technology, under contract with the National Aeronautics and Space Administration. We acknowledge the usage of the HyperLeda database (http://leda.univ-lyon1.fr).

\label{lastpage}


\begin{thebibliography}{}

\bibitem[\protect\citeauthoryear{{Andersson}, {Wannier} \&
  {Morris}}{{Andersson} et~al.}{1991}]{1991ApJ...366..464A}
{Andersson} B.-G.,  {Wannier} P.~G.,    {Morris} M.,  1991, \apj, 366, 464

\bibitem[\protect\citeauthoryear{{Baldry}, {Glazebrook}, {Brinkmann},
  {Ivezi{\'c}}, {Lupton}, {Nichol} \& {Szalay}}{{Baldry}
  et~al.}{2004}]{2004ApJ...600..681B}
{Baldry} I.~K.,  {Glazebrook} K.,  {Brinkmann} J.,  {Ivezi{\'c}} {\v Z}.,
  {Lupton} R.~H.,  {Nichol} R.~C.,    {Szalay} A.~S.,  2004, \apj, 600, 681

\bibitem[\protect\citeauthoryear{{Barnes} \& {Hernquist}}{{Barnes} \&
  {Hernquist}}{1991}]{1991ApJ...370L..65B}
{Barnes} J.~E.,  {Hernquist} L.~E.,  1991, ApJ.L, 370, L65

\bibitem[\protect\citeauthoryear{{Begum}, {Chengalur}, {Kennicutt},
  {Karachentsev} \& {Lee}}{{Begum} et~al.}{2008}]{2008MNRAS.383..809B}
{Begum} A.,  {Chengalur} J.~N.,  {Kennicutt} R.~C.,  {Karachentsev} I.~D.,
  {Lee} J.~C.,  2008, \mnras, 383, 809

\bibitem[\protect\citeauthoryear{{Bell} \& {de Jong}}{{Bell} \& {de
  Jong}}{2001}]{2001ApJ...550..212B}
{Bell} E.~F.,  {de Jong} R.~S.,  2001, \apj, 550, 212

\bibitem[\protect\citeauthoryear{{Bell}, {Wolf}, {Meisenheimer}, {Rix},
  {Borch}, {Dye}, {Kleinheinrich}, {Wisotzki} \& {McIntosh}}{{Bell}
  et~al.}{2004}]{2004ApJ...608..752B}
{Bell} E.~F.,  {Wolf} C.,  {Meisenheimer} K.,  {Rix} H.-W.,  {Borch} A.,  {Dye}
  S.,  {Kleinheinrich} M.,  {Wisotzki} L.,    {McIntosh} D.~H.,  2004, \apj,
  608, 752

\bibitem[\protect\citeauthoryear{{Boissier} \& {Prantzos}}{{Boissier} \&
  {Prantzos}}{1999}]{1999MNRAS.307..857B}
{Boissier} S.,  {Prantzos} N.,  1999, \mnras, 307, 857

\bibitem[\protect\citeauthoryear{{Boselli}}{{Boselli}}{1994}]{1994A&A...292...%
.1B}
{Boselli} A.,  1994, AAP, 292, 1

\bibitem[\protect\citeauthoryear{{Brinchmann}, {Charlot}, {White}, {Tremonti},
  {Kauffmann}, {Heckman} \& {Brinkmann}}{{Brinchmann}
  et~al.}{2004}]{2004MNRAS.351.1151B}
{Brinchmann} J.,  {Charlot} S.,  {White} S.~D.~M.,  {Tremonti} C.,  {Kauffmann}
  G.,  {Heckman} T.,    {Brinkmann} J.,  2004, \mnras, 351, 1151

\bibitem[\protect\citeauthoryear{{Buat} \& {Xu}}{{Buat} \&
  {Xu}}{1996}]{1996A&A...306...61B}
{Buat} V.,  {Xu} C.,  1996, AAP, 306, 61

\bibitem[\protect\citeauthoryear{{Cole}, {Lacey}, {Baugh} \& {Frenk}}{{Cole}
  et~al.}{2000}]{2000MNRAS.319..168C}
{Cole} S.,  {Lacey} C.~G.,  {Baugh} C.~M.,    {Frenk} C.~S.,  2000, \mnras,
  319, 168

\bibitem[\protect\citeauthoryear{{Dale} \& {Helou}}{{Dale} \&
  {Helou}}{2002}]{2002ApJ...576..159D}
{Dale} D.~A.,  {Helou} G.,  2002, Ap.J, 576, 159

\bibitem[\protect\citeauthoryear{{de Vaucouleurs}, {de Vaucouleurs}, {Corwin}
  Jr., {Buta}, {Paturel} \& {Fouque}}{{de Vaucouleurs}
  et~al.}{1991}]{1991trcb.book.....D}
{de Vaucouleurs} G.,  {de Vaucouleurs} A.,  {Corwin} Jr. H.~G.,  {Buta} R.~J.,
  {Paturel} G.,    {Fouque} P.,  1991, {Third Reference Catalogue of Bright
  Galaxies}.
Volume 1-3, XII, 2069 pp.~7 figs..~ Springer-Verlag Berlin Heidelberg New York

\bibitem[\protect\citeauthoryear{{Deharveng}, {Sasseen}, {Buat}, {Bowyer},
  {Lampton} \& {Wu}}{{Deharveng} et~al.}{1994}]{1994A&A...289..715D}
{Deharveng} J.-M.,  {Sasseen} T.~P.,  {Buat} V.,  {Bowyer} S.,  {Lampton} M.,
   {Wu} X.,  1994, AAP, 289, 715

\bibitem[\protect\citeauthoryear{{Dekel} \& {Birnboim}}{{Dekel} \&
  {Birnboim}}{2006}]{2006MNRAS.368....2D}
{Dekel} A.,  {Birnboim} Y.,  2006, \mnras, 368, 2

\bibitem[\protect\citeauthoryear{{Dekel} \& {Silk}}{{Dekel} \&
  {Silk}}{1986}]{1986ApJ...303...39D}
{Dekel} A.,  {Silk} J.,  1986, ApJ, 303, 39

\bibitem[\protect\citeauthoryear{{Federman}, {Glassgold} \& {Kwan}}{{Federman}
  et~al.}{1979}]{1979ApJ...227..466F}
{Federman} S.~R.,  {Glassgold} A.~E.,    {Kwan} J.,  1979, \apj, 227, 466

\bibitem[\protect\citeauthoryear{{Fumagalli}, {Krumholz}, {Prochaska},
  {Gavazzi} \& {Boselli}}{{Fumagalli} et~al.}{2009}]{2009arXiv0903.3950F}
{Fumagalli} M.,  {Krumholz} M.~R.,  {Prochaska} J.~X.,  {Gavazzi} G.,
  {Boselli} A.,  2009, ArXiv e-prints

\bibitem[\protect\citeauthoryear{{Gavazzi}, {Boselli}, {Donati}, {Franzetti} \&
  {Scodeggio}}{{Gavazzi} et~al.}{2003}]{2003A&A...400..451G}
{Gavazzi} G.,  {Boselli} A.,  {Donati} A.,  {Franzetti} P.,    {Scodeggio} M.,
  2003, AAP, 400, 451

\bibitem[\protect\citeauthoryear{{Gerola}, {Seiden} \& {Schulman}}{{Gerola}
  et~al.}{1980}]{1980ApJ...242..517G}
{Gerola} H.,  {Seiden} P.~E.,    {Schulman} L.~S.,  1980, ApJ, 242, 517

\bibitem[\protect\citeauthoryear{{Hernquist}}{{Hernquist}}{1989}]{1989Natur.34%
0..687H}
{Hernquist} L.,  1989, Nat, 340, 687

\bibitem[\protect\citeauthoryear{{Holmberg}}{{Holmberg}}{1958}]{1958MeLu2.136.%
...1H}
{Holmberg} E.,  1958, Meddelanden fran Lunds Astronomiska Observatorium Serie
  II, 136, 1

\bibitem[\protect\citeauthoryear{{Hopkins}, {Hernquist}, {Cox}, {Di Matteo},
  {Robertson} \& {Springel}}{{Hopkins} et~al.}{2006}]{2006ApJS..163....1H}
{Hopkins} P.~F.,  {Hernquist} L.,  {Cox} T.~J.,  {Di Matteo} T.,  {Robertson}
  B.,    {Springel} V.,  2006, ApJ.S, 163, 1

\bibitem[\protect\citeauthoryear{{Hunter} \& {Elmegreen}}{{Hunter} \&
  {Elmegreen}}{2004}]{2004AJ....128.2170H}
{Hunter} D.~A.,  {Elmegreen} B.~G.,  2004, A.J., 128, 2170

\bibitem[\protect\citeauthoryear{{Hunter}, {Elmegreen} \& {Baker}}{{Hunter}
  et~al.}{1998}]{1998ApJ...493..595H}
{Hunter} D.~A.,  {Elmegreen} B.~G.,    {Baker} A.~L.,  1998, \apj, 493, 595

\bibitem[\protect\citeauthoryear{{James}, {Shane}, {Beckman}, {Cardwell},
  {Collins}, {Etherton}, {de Jong}, {Fathi}, {Knapen}, {Peletier}, {Percival},
  {Pollacco}, {Seigar}, {Stedman} \& {Steele}}{{James}
  et~al.}{2004}]{2004A&A...414...23J}
{James} P.~A.,  {Shane} N.~S.,  {Beckman} J.~E.,  {Cardwell} A.,  {Collins}
  C.~A.,  {Etherton} J.,  {de Jong} R.~S.,  {Fathi} K.,  {Knapen} J.~H.,
  {Peletier} R.~F.,  {Percival} S.~M.,  {Pollacco} D.~L.,  {Seigar} M.~S.,
  {Stedman} S.,    {Steele} I.~A.,  2004, AAP, 414, 23

\bibitem[\protect\citeauthoryear{{Kaisin}, {Kasparova}, {Knyazev} \&
  {Karachentsev}}{{Kaisin} et~al.}{2007}]{2007AstL...33..283K}
{Kaisin} S.~S.,  {Kasparova} A.~V.,  {Knyazev} A.~Y.,    {Karachentsev} I.~D.,
  2007, Astronomy Letters, 33, 283

\bibitem[\protect\citeauthoryear{{Karachentsev}, {Karachentseva}, {Huchtmeier}
  \& {Makarov}}{{Karachentsev} et~al.}{2004}]{2004AJ....127.2031K}
{Karachentsev} I.~D.,  {Karachentseva} V.~E.,  {Huchtmeier} W.~K.,    {Makarov}
  D.~I.,  2004, AJ, 127, 2031

\bibitem[\protect\citeauthoryear{{Kenney}}{{Kenney}}{1997}]{1997ASSL..161...33%
K}
{Kenney} J.,  1997, in Astrophysics and Space Science Library Vol.~161 of
  Astrophysics and Space Science Library, {Molecular gas in galaxy disks.}.
pp 33--74

\bibitem[\protect\citeauthoryear{{Kenney} \& {Young}}{{Kenney} \&
  {Young}}{1989}]{1989ApJ...344..171K}
{Kenney} J.~D.~P.,  {Young} J.~S.,  1989, \apj, 344, 171

\bibitem[\protect\citeauthoryear{{Kennicutt}
  Jr.}{{Kennicutt}}{1983}]{1983ApJ...272...54K}
{Kennicutt} Jr. R.~C.,  1983, \apj, 272, 54

\bibitem[\protect\citeauthoryear{{Kennicutt}
  Jr.}{{Kennicutt}}{1998a}]{1998ARA&A..36..189K}
{Kennicutt} Jr. R.~C.,  1998a, ARAA, 36, 189

\bibitem[\protect\citeauthoryear{{Kennicutt}
  Jr.}{{Kennicutt}}{1998b}]{1998ApJ...498..541K}
{Kennicutt} Jr. R.~C.,  1998b, \apj, 498, 541

\bibitem[\protect\citeauthoryear{{Kennicutt} Jr. et~al.,}{{Kennicutt}
  et~al.}{2009}]{Kennicutt_prep_09}
{Kennicutt} Jr. R.~C.,  et~al., 2009, ApJ submitted

\bibitem[\protect\citeauthoryear{{Kennicutt} Jr., {Lee}, {Funes} Jos{\'e}~G.,
  {Sakai} \& {Akiyama}}{{Kennicutt} et~al.}{2008}]{2008ApJS..178..247K}
{Kennicutt} Jr. R.~C.,  {Lee} J.~C.,  {Funes} Jos{\'e}~G. S.~J.,  {Sakai} S.,
   {Akiyama} S.,  2008, \apjs, 178, 247

\bibitem[\protect\citeauthoryear{{Kroupa}, {Tout} \& {Gilmore}}{{Kroupa}
  et~al.}{1993}]{1993MNRAS.262..545K}
{Kroupa} P.,  {Tout} C.~A.,    {Gilmore} G.,  1993, \mnras, 262, 545

\bibitem[\protect\citeauthoryear{{Krumholz}, {McKee} \& {Tumlinson}}{{Krumholz}
  et~al.}{2009}]{2009ApJ...699..850K}
{Krumholz} M.~R.,  {McKee} C.~F.,    {Tumlinson} J.,  2009, \apj, 699, 850

\bibitem[\protect\citeauthoryear{{Lee}, {Gil de Paz}, {Tremonti}, {Kennicutt}
  et~al.,}{{Lee} et~al.}{2009}]{Lee_09}
{Lee} J.~C.,  {Gil de Paz} A.,  {Tremonti} C.,  {Kennicutt} R.~C.,    et~al.,
  2009, ApJ submitted

\bibitem[\protect\citeauthoryear{{Lee}, {Kennicutt}, {Funes} Jos{\'e}~G.,
  {Sakai} \& {Akiyama}}{{Lee} et~al.}{2007}]{2007ApJ...671L.113L}
{Lee} J.~C.,  {Kennicutt} R.~C.,  {Funes} Jos{\'e}~G. S.~J.,  {Sakai} S.,
  {Akiyama} S.,  2007, \apjl, 671, L113

\bibitem[\protect\citeauthoryear{{Lee}, {Kennicutt}, {Jos{\'e} G.~Funes},
  {Sakai} \& {Akiyama}}{{Lee} et~al.}{2009}]{2009ApJ...692.1305L}
{Lee} J.~C.,  {Kennicutt} R.~C.,  {Jos{\'e} G.~Funes} S.~J.,  {Sakai} S.,
  {Akiyama} S.,  2009, \apj, 692, 1305

\bibitem[\protect\citeauthoryear{{Leroy}, {Bolatto}, {Simon} \&
  {Blitz}}{{Leroy} et~al.}{2005}]{2005ApJ...625..763L}
{Leroy} A.,  {Bolatto} A.~D.,  {Simon} J.~D.,    {Blitz} L.,  2005, \apj, 625,
  763

\bibitem[\protect\citeauthoryear{{Mac Low} \& {Ferrara}}{{Mac Low} \&
  {Ferrara}}{1999}]{1999ApJ...513..142M}
{Mac Low} M.-M.,  {Ferrara} A.,  1999, ApJ, 513, 142

\bibitem[\protect\citeauthoryear{{Morris} \& {Rickard}}{{Morris} \&
  {Rickard}}{1982}]{1982ARA&A..20..517M}
{Morris} M.,  {Rickard} L.~J.,  1982, \araa, 20, 517

\bibitem[\protect\citeauthoryear{{Moustakas} \& {Kennicutt} Jr.}{{Moustakas} \&
  {Kennicutt}}{2006}]{2006ApJS..164...81M}
{Moustakas} J.,  {Kennicutt} Jr. R.~C.,  2006, \apjs, 164, 81

\bibitem[\protect\citeauthoryear{{Noeske}, {Faber}, {Weiner}, {Koo}, {Primack},
  {Dekel}, {Papovich}, {Conselice}, {Le Floc'h}, {Rieke}, {Coil}, {Lotz},
  {Somerville} \& {Bundy}}{{Noeske} et~al.}{2007}]{2007ApJ...660L..47N}
{Noeske} K.~G.,  {Faber} S.~M.,  {Weiner} B.~J.,  {Koo} D.~C.,  {Primack}
  J.~R.,  {Dekel} A.,  {Papovich} C.,  {Conselice} C.~J.,  {Le Floc'h} E.,
  {Rieke} G.~H.,  {Coil} A.~L.,  {Lotz} J.~M.,  {Somerville} R.~S.,    {Bundy}
  K.,  2007, \apjl, 660, L47

\bibitem[\protect\citeauthoryear{{Obreschkow} \& {Rawlings}}{{Obreschkow} \&
  {Rawlings}}{2009}]{2009arXiv0901.2526O}
{Obreschkow} D.,  {Rawlings} S.,  2009, ArXiv e-prints

\bibitem[\protect\citeauthoryear{{Paturel}, {Petit}, {Prugniel}, {Theureau},
  {Rousseau}, {Brouty}, {Dubois} \& {Cambr{\'e}sy}}{{Paturel}
  et~al.}{2003}]{2003A&A...412...45P}
{Paturel} G.,  {Petit} C.,  {Prugniel} P.,  {Theureau} G.,  {Rousseau} J.,
  {Brouty} M.,  {Dubois} P.,    {Cambr{\'e}sy} L.,  2003, AAP, 412, 45

\bibitem[\protect\citeauthoryear{{Pflamm-Altenburg}, {Weidner} \&
  {Kroupa}}{{Pflamm-Altenburg} et~al.}{2007}]{2007ApJ...671.1550P}
{Pflamm-Altenburg} J.,  {Weidner} C.,    {Kroupa} P.,  2007, ApJ, 671, 1550

\bibitem[\protect\citeauthoryear{{Roberts}}{{Roberts}}{1969}]{1969AJ.....74..8%
59R}
{Roberts} M.~S.,  1969, \aj, 74, 859

\bibitem[\protect\citeauthoryear{{Roberts} \& {Haynes}}{{Roberts} \&
  {Haynes}}{1994}]{1994ARA&A..32..115R}
{Roberts} M.~S.,  {Haynes} M.~P.,  1994, ARAA, 32, 115

\bibitem[\protect\citeauthoryear{{Salim}, {Rich}, {Charlot}, {Brinchmann},
  {Johnson} \& {Schiminovich}}{{Salim} et~al.}{2007}]{2007tS..173..267S}
{Salim} S.,  {Rich} R.~M.,  {Charlot} S.,  {Brinchmann} J.,  {Johnson} B.~D.,
   {Schiminovich} D.,  2007, Ap.JS, 173, 267

\bibitem[\protect\citeauthoryear{{Sandage}}{{Sandage}}{1986}]{1986A&A...161...%
89S}
{Sandage} A.,  1986, \aap, 161, 89

\bibitem[\protect\citeauthoryear{{Scalo}}{{Scalo}}{1986}]{1986FCPh...11....1S}
{Scalo} J.~M.,  1986, Fundamentals of Cosmic Physics, 11, 1

\bibitem[\protect\citeauthoryear{{Schaye}}{{Schaye}}{2004}]{2004ApJ...609..667%
S}
{Schaye} J.,  2004, \apj, 609, 667

\bibitem[\protect\citeauthoryear{{Schiminovich}, {Wyder}, {Martin} \&
  {Johnson}}{{Schiminovich} et~al.}{2007}]{2007ApJS..173..315S}
{Schiminovich} D.,  {Wyder} T.~K.,  {Martin} D.~C.,    {Johnson} B.~D.,  2007,
  \apjs, 173, 315

\bibitem[\protect\citeauthoryear{{Schlegel}, {Finkbeiner} \&
  {Davis}}{{Schlegel} et~al.}{1998}]{1998ApJ...500..525S}
{Schlegel} D.~J.,  {Finkbeiner} D.~P.,    {Davis} M.,  1998, Ap.J, 500, 525

\bibitem[\protect\citeauthoryear{{Schmidt}}{{Schmidt}}{1959}]{1959ApJ...129..2%
43S}
{Schmidt} M.,  1959, \apj, 129, 243

\bibitem[\protect\citeauthoryear{{Searle}, {Sargent} \& {Bagnuolo}}{{Searle}
  et~al.}{1973}]{1973ApJ...179..427S}
{Searle} L.,  {Sargent} W.~L.~W.,    {Bagnuolo} W.~G.,  1973, ApJ, 179, 427

\bibitem[\protect\citeauthoryear{{Shaya} \& {Federman}}{{Shaya} \&
  {Federman}}{1987}]{1987ApJ...319...76S}
{Shaya} E.~J.,  {Federman} S.~R.,  1987, \apj, 319, 76

\bibitem[\protect\citeauthoryear{{Spergel}, {Bean}, {Dor{\'e}}, {Nolta},
  {Bennett}, {Dunkley}, {Hinshaw}, {Jarosik}, {Komatsu}, {Page} \&
  {Peiris}}{{Spergel} et~al.}{2007}]{2007ApJS..170..377S}
{Spergel} D.~N.,  {Bean} R.,  {Dor{\'e}} O.,  {Nolta} M.~R.,  {Bennett} C.~L.,
  {Dunkley} J.,  {Hinshaw} G.,  {Jarosik} N.,  {Komatsu} E.,  {Page} L.,
  {Peiris} H.~V.,  2007, \apjs, 170, 377

\bibitem[\protect\citeauthoryear{{Springob}, {Haynes}, {Giovanelli} \&
  {Kent}}{{Springob} et~al.}{2005}]{2005ApJS..160..149S}
{Springob} C.~M.,  {Haynes} M.~P.,  {Giovanelli} R.,    {Kent} B.~R.,  2005,
  \apjs, 160, 149

\bibitem[\protect\citeauthoryear{{Stinson}, {Dalcanton}, {Quinn}, {Kaufmann} \&
  {Wadsley}}{{Stinson} et~al.}{2007}]{2007ApJ...667..170S}
{Stinson} G.~S.,  {Dalcanton} J.~J.,  {Quinn} T.,  {Kaufmann} T.,    {Wadsley}
  J.,  2007, ApJ, 667, 170

\bibitem[\protect\citeauthoryear{{Strateva}, {Ivezi{\'c}}, {Knapp}, {Narayanan}
  \& {Strauss}}{{Strateva} et~al.}{2001}]{2001AJ....122.1861S}
{Strateva} I.,  {Ivezi{\'c}} {\v Z}.,  {Knapp} G.~R.,  {Narayanan} V.~K.,
  {Strauss} M.~A.,  2001, AJ, 122, 1861

\bibitem[\protect\citeauthoryear{{Tremonti}, {Heckman}, {Kauffmann},
  {Brinchmann}, {Charlot}, {White}, {Seibert}, {Peng}, {Schlegel}, {Uomoto},
  {Fukugita} \& {Brinkmann}}{{Tremonti} et~al.}{2004}]{2004ApJ...613..898T}
{Tremonti} C.~A.,  {Heckman} T.~M.,  {Kauffmann} G.,  {Brinchmann} J.,
  {Charlot} S.,  {White} S.~D.~M.,  {Seibert} M.,  {Peng} E.~W.,  {Schlegel}
  D.~J.,  {Uomoto} A.,  {Fukugita} M.,    {Brinkmann} J.,  2004, \apj, 613, 898

\bibitem[\protect\citeauthoryear{{Wong} \& {Blitz}}{{Wong} \&
  {Blitz}}{2002}]{2002ApJ...569..157W}
{Wong} T.,  {Blitz} L.,  2002, \apj, 569, 157

\bibitem[\protect\citeauthoryear{{Young} \& {Scoville}}{{Young} \&
  {Scoville}}{1991}]{1991ARA&A..29..581Y}
{Young} J.~S.,  {Scoville} N.~Z.,  1991, ARAA, 29, 581

\bibitem[\protect\citeauthoryear{{Young}, {Xie}, {Tacconi}, {Knezek},
  {Viscuso}, {Tacconi-Garman}, {Scoville}, {Schneider}, {Schloerb}, {Lord},
  {Lesser}, {Kenney}, {Huang}, {Devereux}, {Claussen}, {Case}, {Carpenter},
  {Berry} \& {Allen}}{{Young} et~al.}{1995}]{1995ApJS...98..219Y}
{Young} J.~S.,  {Xie} S.,  {Tacconi} L.,  {Knezek} P.,  {Viscuso} P.,
  {Tacconi-Garman} L.,  {Scoville} N.,  {Schneider} S.,  {Schloerb} F.~P.,
  {Lord} S.,  {Lesser} A.,  {Kenney} J.,  {Huang} Y.-L.,  {Devereux} N.,
  {Claussen} M.,  {Case} J.,  {Carpenter} J.,  {Berry} M.,    {Allen} L.,
  1995, \apjs, 98, 219

\end{thebibliography}
\end{document}